\documentclass[a4paper,11pt]{article}
\usepackage{aaskaiid}
\usepackage{graphicx, amsmath, caption, subcaption, isomath, bm} % Required for inserting images
\usepackage{newtxtext,newtxmath, color, xcolor}
\usepackage[pagewise]{lineno}
\usepackage[normalem]{ulem}
\usepackage[flushleft]{threeparttable}
\usepackage{pdflscape}
\usepackage{afterpage}
\usepackage{capt-of}% or use the larger `caption` package
\usepackage{lipsum}% dummy text
\usepackage[utf8]{inputenc}
\usepackage{orcidlink}
\def\visbnut{\mathcal{V}\left( \mathbfit{b}, \nu, t \right)}
\def\visbnutm{\mathcal{V}\left( \mathbfit{b}, \nu_m, t_m \right)}
\def\r{\hat{\mathbfit{r}}}

\def\rcm{\hat{\mathbfit{r}}_c(t_m)}
\def\Irnut{\mathcal{I}\left( \r, \nu, t \right)}
\def\Brnut{\mathcal{B}\left( \r, \nu, t \right)}
\def\Irnutm{\mathcal{I}\left( \r, \nu_m, t_m \right)}
\def\Brnutm{\mathcal{B}\left( \r, \nu_m, t_m \right)}
\def\bt{\mathbfit{b}(t)}
\def\btm{\mathbfit{b}(t_m)}
\def\dbdtm{\frac{d\mathbfit{b}}{dt}\vert_{t=t_m}}

\def\nubc{\frac{\nu}{c}}
\def\numbc{\frac{\nu_m}{c}}
\def\dt{\delta t}
\def\dnu{\delta \nu}
\def\sinc{{\rm sinc}}

\def\eg{{\it e.g.}\,}

\newcommand{\HI}{{\rm H\hspace{0.5mm}}{\scriptsize {\rm I}}}

\title{Fast Simultaneous Surveys with On-the-Fly Mapping }
\ShortTitle{On-the-Fly}

\author[1]{Suman Chatterjee$^*$\orcidlink{0000-0001-8852-5888}}
\emailAdd{$^*$sumanchttrj05@gmail.com, 6306414@myuwc.ac.za}
\author[2]{Sarvesh Mangla$^{**}$\orcidlink{0000-0003-3485-5122}}\emailAdd{\\$^{**}$ mangla.sarvesh@physik.lmu.de}
\author[3]{Sourabh Paul$^{***}$\orcidlink{0000-0002-8671-2177}}\emailAdd{$^{***}$sourabh.paul@manchester.ac.uk}
\author[3]{Keith Grainge\orcidlink{0000-0002-6780-1406}}
\author[4]{Matthias Hoeft\orcidlink{0000-0001-5571-1369}}
\author[1,5]{Tamera Kassie \orcidlink{0000-0002-3965-0057}}
\author[2]{Joseph J. Mohr \orcidlink{0000-0002-6875-2087}}
\author[6]{Yvette Perrott \orcidlink{0000-0002-6255-8240}}
\author[2]{Kristof Rozgonyi \orcidlink{0000-0002-5611-9292}}
\author[1, 5]{Mario G. Santos\orcidlink{0000-0003-3892-3073}}
\author[5, 7, 8]{Oleg M. Smirnov \orcidlink{0000-0003-1680-7936}}
\author[9, 10]{Cyril Tasse \orcidlink{0009-0009-9030-7885}}
\author[3]{Laura Wolz\orcidlink{0000-0003-3334-3037}}

\affiliation[1]{Department of Physics and Astronomy, University of the Western Cape, Robert Sobukwe Road, Bellville, Cape Town 7535, South Africa}
\affiliation[2]{University Observatory, Faculty of Physics,  Ludwig-Maximilians-Universität, Scheinerstr. 1, 81679, Munich, Germany}
\affiliation[3]{Jodrell Bank Centre for Astrophysics, Department of Physics \& Astronomy, The University of Manchester, Manchester M13 9PL, UK}
\affiliation[4]{Thüringer Landessternwarte, Sternwarte 5, 07778 Tautenburg, Germany}
\affiliation[5]{South African Radio Astronomy Observatory (SARAO), Liesbeek House, River Park, Gloucester Road, Mowbray, Cape Town, 7700, South Africa}
\affiliation[6]{School of Chemical and Physical Sciences, Victoria University of Wellington, Wellington 6012, New Zealand}
\affiliation[7]{Centre for Radio Astronomy Techniques and Technologies (RATT), Department of Physics and Electronics, Rhodes University, Makhanda, 6140, South Africa}

\affiliation[8]{Institute for Radioastronomy, National Institute of Astrophysics (INAF IRA), Via Gobetti 101, 40129 Bologna, Italy}
\affiliation[9]{GEPI \& ORN, Observatoire de Paris, Université PSL, CNRS, 5 Place Jules Janssen, 92190 Meudon, France}
\affiliation[10]{Department of Physics \& Electronics, Rhodes University, PO Box 94, Grahamstown, 6140, South Africa}

\abstract{The SKAO is a next-generation radio telescope that will transform our understanding of the formation and evolution of radio galaxies, quasars, transients, and other cosmic sources. Surveys conducted on precursor telescopes can inform us about the capabilities and challenges to be overcome in preparation for SKAO science. The MeerKAT Large Area Synoptic Survey (MeerKLASS), is a pathfinder large area survey to probe cosmology using the single-dish \HI~intensity mapping (IM). MeerKLASS provides an additional wide, high angular-resolution commensal survey by utilizing the ``On-the-Fly'' (OTF) imaging technique. The survey target is to cover 10,000 sq. degrees in the Southern sky avoiding the Galactic plane, using the UHF-band (544-1088 MHz). The survey aims to achieve an r.m.s. of 25 $\mu$Jy/beam and $14''$ resolution. In this chapter we discuss the survey strategy, observational status, the development of the MeerKLASS OTF imaging pipeline and the science prospects of the data products. OTF imaging, which relies upon constant elevation fast scanning, comes with its own set of challenges, such as smearing. Here we detail how we have mitigated them and the current scientific impact. Significantly fast survey speed ($\sim 3.5$ sq. deg per min), commensality with \HI~IM survey, deep and large area continuum images make MeerKLASS OTF an economic survey technique. Lessons from this technique will be valuable for the upcoming SKA-Mid, where we expect a better resolution $(< 2'')$ and sensitivity $(\sim 7\mu {\rm Jy/beam})$ of the OTF images with the AA4 configuration. Furthermore, improvements in the correlator will result in negligible smearing effects in the imaging.}

\begin{document}
\maketitle

\section{Introduction}
Over the past decade, we have seen multiple large area radio surveys from very low frequencies (\eg 70 MHz, \citealt{LOLSS_2023}) to as high as 20 GHz \citep{Murphy2010}. It is also becoming evident that the impact of these wide-area radio surveys are undeniable in the field of cosmology, galaxy cluster and large-scale structures in the universe \citep{Alonso2021, Hale2023}. These large-area surveys are playing a crucial role in understanding the evolution of galaxies, properties of nearby galaxies and magnetic sky. As we start observing with the SKA-Mid and Low, then such wide area surveys are expected to be much deeper. 

Before SKA, several of its pathfinder instruments have invested in precursor synoptic surveys to characterise stellar and galaxy population to unprecedented detail \citep{Hurley-Walker2017, McConnell2020, Hopkins_2025, Hale01.2026.SKA}. A description of the surveys is provided in \autoref{tab:survey_SH}. However, these surveys usually take many years of observations to complete, primarily due to the point-and-track observations strategy. These traditional observational strategies have a significant overhead and a smaller footprint which leads to slow survey speed. Furthermore, different science cases require different amount of resources, for example, neutral hydrogen (\HI) intensity mapping (IM) experiments require long observations to beat down noise and other systematics  \citep{Spinelli02.2026.SKA}, whereas transients surveys require covering a vary large area of sky repeatedly, requiring fast scanning.

To achieve an elevated efficiency of a wide-area multi-epoch survey, it is necessary to rapidly slew the telescope across the sky. The observed data is then collected continuously at a regular interval during this rapid slew of the telescope in what is called `` On-The-Fly'' (OTF; \citealt{Mangum2007, Sawada2008, Mooley2019, Chatterjee2025}). The OTF also uses the fact that the next generation telescopes, such as SKA, will have large instantaneous sensitivity. The main scientific aims of a generic wide-area radio survey are to explore the evolution of galaxies, clusters and the large scale structures, search for slow transients and measure magnetic fields through rotation measure synthesis of polarisation data.

In this chapter, we describe the OTF capabilities of SKA-Mid pathfinder MeerKAT \footnote{\href{https://www.sarao.ac.za/science/meerkat/about-meerkat/}{https://www.sarao.ac.za/science/meerkat/}} that takes advantage of highly sensitive, low-noise receivers and also show the importance of commensal observation planning.  MeerKAT  consists of $64 \times 13.5$-m dishes with offset Gregorian optics, providing an unblocked aperture. It is equipped with three receiver bands: UHF (544-1088 MHz), L band (856-1712 MHz), and S band (1750–3500 MHz). Three-quarters of the collecting area is within a dense, 1 km diameter core region, and the remaining dishes are situated around the core, providing a maximum baseline of 8 km. The large number of baselines, wide field of view (2 deg at UHF-band), and low ($\sim 20$ K) system temperature all combine to make MeerKAT an exceptionally fast and capable synthesis imaging telescope. The correlator can also deliver up to $32,768$ frequency channels, resulting in excellent spectroscopic imaging capabilities and capability for excellent RFI excision. 

The MeerKAT Large Area Synoptic Survey (MeerKLASS; \citealt{Santos2017}) plans to observe $\sim 10,000$ sq. degrees in the Southern sky outside the galactic plane over the next few years. 
The survey is performed both in the auto-correlation (single dish mode) and the cross-correlation (interferometric mode). With the single dish observations, the survey aims to perform the \HI~IM up to redshift of $z = 1.44$. MeerKLASS auto-correlations survey aims for a statistical detection of the \HI~IM signal both in auto-power spectrum and cross-correlation with other surveys \citep{Cunnington2022, Cunnington2025, Cunnington2025b, Cunnington01.2026.SKA}.  In this chapter, we discuss a commensal OTF continuum survey with the interferometric visibilities recorded during MeerKLASS observations. The MeerKLASS target sky area has significant overlap with several optical/NIR,  wide, galaxy surveys and is expected to provide an invaluable legacy data set.

% This will allow us to measure  Baryon Acoustic Oscillations (BAO) and redshift-space distortion (RSD), in turn enabling us to constrain different dark energy and modified gravity theories. The large scale \HI~IM survey can be used to constrain the primordial non-gaussianity \citep{Camera2013, Alonso_2015}. Although the majority of the MeerKLASS observation is planned to be done in MeerKAT UHF band, we have also performed several observations in MeerKAT L-band such as $200\,{\rm deg}^2$ around WiggleZ 11-h field \citep{Wang2021, Cunnington2022}, $236\,{\rm deg}^2$ around KiDS-South field \citep{Cunnington2025}. In the UHF band we observed approximately $2,500\,{\rm deg}^2$ sky area that has overlap with DESI-Y1 observations \citep{DESI_Y1-2025}. 

197 dishes of future SKA-Mid are arranged in a tightly packed central core with three spiral arms extending outward, providing baselines of up to 150 km \citep{Braun2019}. These long separations are crucial for achieving high angular resolution interferometric observations. At its completion, the array will have a total collecting area of 33,000 sq. meters and cover frequencies from 350 MHz to 15.4 GHz, with an aspirational upper limit of 24 GHz. In this chapter, we focus on Band 1 of SKA-Mid that will operate between a frequency range 0.35-1.05 GHz. Our aim here is to explore the possibility of OTF mapping similar to MeerKLASS OTF survey using Band 1 of SKA-Mid.

The chapter is organized as follows. In \autoref{sec:motf} we discuss the basic survey strategy implemented for MerrKLASS OTF survey, challenges and the impact of that in data products. \autoref{sec:pipe} describes the pipeline built for OTF mapping. In \autoref{sec:lband} and \autoref{sec:uhfband} we discuss what have been achieved using L and UHF-band of the MerrKAT, respectively. We present our OTF-mapping forecast for SKA-Mid in \autoref{sec:ska}, and conclude the chapter with a summary of key points in \autoref{sec:summary}.

\begin{table*}
\caption{A non-exhaustive summary of recent large area low-frequency surveys (see also \autoref{fig:comp_sensitivity}). We have attempted to provide a fair comparison of sensitivities and resolutions, but we note that both the sensitivity and resolution achieved vary within a given survey.}
 \centering
 \label{tab:survey_SH}
\begin{tabular}{lcccccc}
\hline 
Survey & Resolution & Noise & Frequency  & Declination & Area \\
                            & ($''$) & ($\mu$Jy/beam) & (MHz) &  & (${\rm degree}^{2}$)\\ \hline
TGSS ADR  & 25  & 3500            & 140--156 & $\delta>-53^\circ$ & 36,900\\
(\citealt{Intema2016}) &   &             &  &  & \\
GLEAM & 150  & 5000 & 72--231 & $\delta<+25^\circ$ & 24,831\\
(\citealt{Wayth2015}) &   &             &  &  & \\

LOTSS DR2 & 6  & 83 & 120--168 & $\delta>+15^\circ$ & 5,635\\
(\citealt{Shimwell2022}) &   &             &  &  & \\
RACS-Low  & 18 & 260           & 743.5--1031.5 & $\delta<+30^\circ$ & 30,480\\
(\citealt{Hale2021}) &   &             &  &  & \\
RACS-Mid  & 11.2 & 140           & 1151.5--1439.5 & $\delta<+49^\circ$ & 36,200\\
(\citealt{Duchesne2023}) &   &             &  &  & \\
RACS-High  & 11.8 & 195           & 1511.5--1799.5 & $\delta<+48^\circ$ & 37,000 \\
(\citealt{duchesne2025}) &   &             &  &  & \\
EMU  & 15 & 20-30           & 800--1088 & $\delta<+30^\circ$ & 20,626\\
(\citealt{Norris2011}, &   &             &  &  & \\
\citealt{Hopkins_2025}) &   &             &  &  & \\
MALS  & 7.7 & 10           & 856--1712 & $\delta<+30^\circ$ & 4,344\\
(\citealt{Gupta2016}, &   &             &  &  & \\
\citealt{Wagenveld2024}) &   &             &  &  & \\
SUMSS  & 45 & 1500           & 843 & $\delta<-30^\circ$ & 8,000\\
(\citealt{Mauch2003}) &   &             &  &  & \\
NVSS  & 45 & 450           & 1400 & $\delta>-40^\circ$ &33,000\\
(\citealt{Condon1998})&   &             &  &  & \\
VLASS  & 2.5 & 70           & 2000 -- 4000 & $\delta>-40^\circ$ &33,885\\
(\citealt{Lacy2020}) &   &             &  &  & \\
MeerKLASS (UHF) & $14[23.3]^a$  & 25 & 580--1015 & $\delta<+10^\circ$ &10,000\\
(\citealt{Paul2025}) &   &             &  &  & \\
MeerKLASS (L) & $11[14]^b$  & 20 & 856--1712 & $\delta<+10^\circ$ & 500\\ 
(\citealt{Mangla2025}) &   &             &  &  & \\\hline
% LoTSS direction-dependent  & 5 & 0.1 & 120--168 & $\delta>0^\circ$ \\ 
% LoTSS direction-independent (this paper)  & 25 & 0.5 & 120--168 & HETDEX Spring Field \\ 
% VLSSr (\citealt{Lane_2014}) & 75     & 100           &  73--74.6 & $\delta>-30^\circ$ \\ \hline
 \end{tabular}

 %##################
\begin{tablenotes}[flushleft]
	\item \footnotesize{a: The target resolution is $14''$, whereas `[]' shows the average beam after the smearing correction, where ${\rm beam} = \sqrt{b_{\rm maj}b_{\rm min}} = \sqrt{32'' \times 17''}$.}
	\item \footnotesize{b: Same as  mentioned in `a', where ${\rm beam}  = \sqrt{25'' \times 8''}$.}
\end{tablenotes}
%################## 
\end{table*}
\begin{figure}
    \centering
    \includegraphics[width=1.0\linewidth]{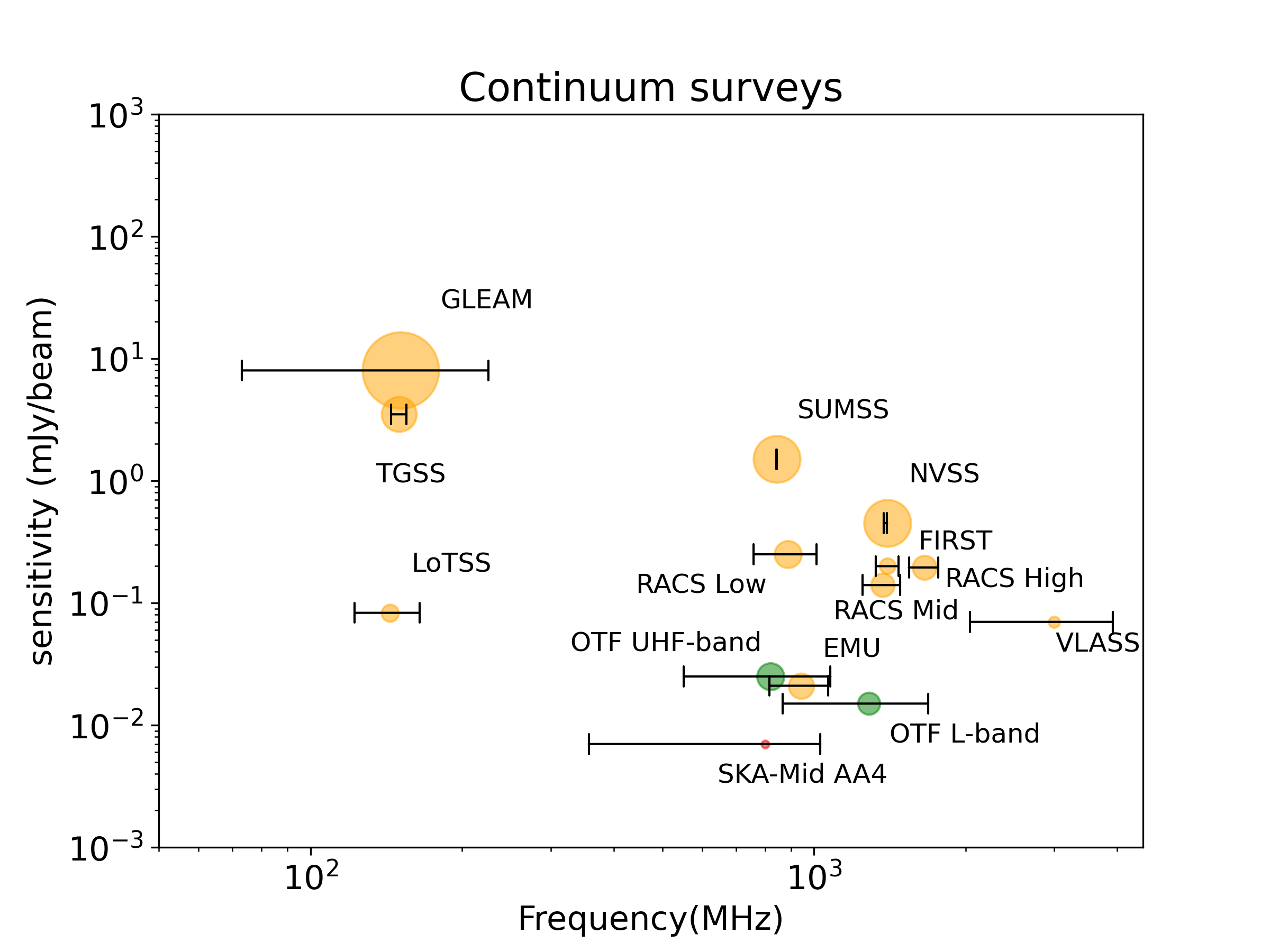}
    \caption{A summary of recent large area low-frequency surveys in including their sensitivity, frequency, and resolution (see Table~\ref{tab:survey_SH}). The size of the markers is proportional to the square root of the survey resolution.  The horizontal lines show the frequency coverage for surveys. The green circles show the expected sensitivity and resolution of MeerKLASS OTF surveys without smearing incorporated. Red circle show the expected resolution and the sensitivity achievable using OTF mapping with SKA-Mid AA4 configuration.}
    \label{fig:comp_sensitivity}
\end{figure}

\section{MeerKLASS OTF formalism} \label{sec:motf}
The MeerKLASS scanning strategy is optimised for the single-dish mode \citep{Wang2021}, however, visibilities are recorded for all scans and using the OTF mapping technique, continuum images can be produced from all data. For the rest of this chapter, we will refer to the MeerKLASS OTF data products as ``M-OTF''. Each M-OTF observation is made with the full instantaneous bandwidth of the telescope divided into 4096 contiguous frequency channels and with a sampling  of $\dt =2{\rm s}$. All four polarization of the visibilities (XX, XY, YX, YY) are recorded to allow images to be made in stokes parameters I, Q, U and V. 

% The total M-OTF survey is expected to conservatively achieve a flux sensitivity of $25-35\mu{\rm Jy}$ in stokes I and an angular resolution of 14'' (robust weighted 0) over $10,000\,{\rm deg}^2$. Table~\ref{tab:survey_SH} shows the expected sensitivity of the survey in the Southern Sky and figure~\ref{fig:comp_sensitivity} describes the same.
\subsection{Survey Strategy}
\begin{figure}
    \centering
    \includegraphics[width=.9\linewidth]{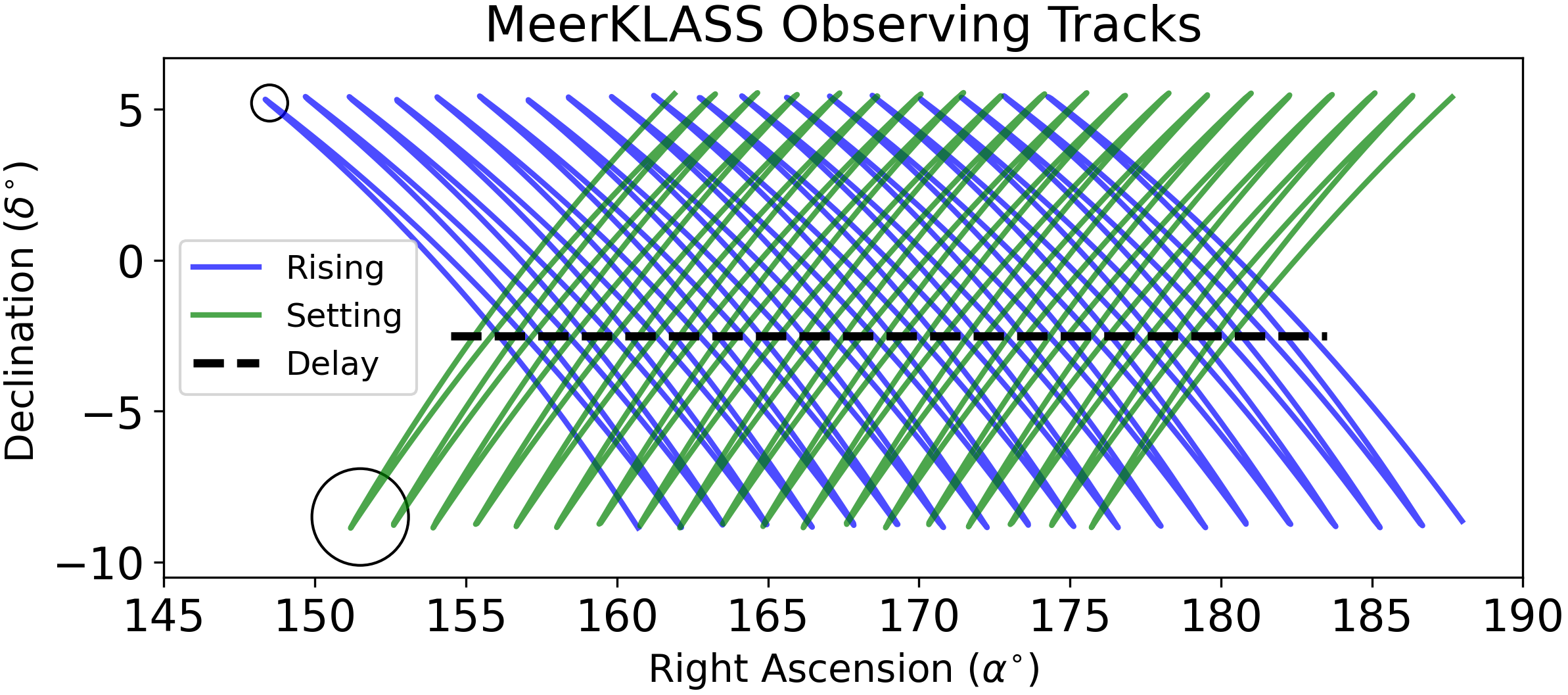}
    \caption{An example of MeerKLASS constant elevation observing strategy where ``rising" and ``setting" epochs are shown in blue and green, respectively. L and UHF-band primary beam FWHM at the nominal frequencies are shown using circles. Each epoch lasts approximately 100 mins. This region is called a `box' according to our naming conversion. The horizontal dashed line passing through the center of the scanning pattern marks the sky region where the delay for the visibilities are set while the dishes perform fast scanning.}
    \label{fig:scan}
\end{figure}
A constant elevation fast scanning strategy is adopted to optimise the coverage of the relevant large angular scales and the stability time-scale of the instrument. MeerKAT dishes are moved back and forth to scan in the azimuth (az) direction at a constant elevation (el) with a slew of $\sim 10\ {\rm deg}$ in the sky (Fig.~\ref{fig:scan}). This also minimizes fluctuations in the ground spill and air-mass \citep{Wang2021}. One end-to-end slew is named `scan' in the subsequent text. The telescope speed for a scan is set to  $\Theta = 7/\cos(el) \,{\rm arcmin \, s}^{-1}$ so that the projected scan speed on the sky is fixed a $7 \,{\rm arcmin \, s}^{-1}$ which ensures that the telescope pointing does not shift significantly compared to the full-width-half-maximum (FWHM) of the primary beam (at ${\rm UHF - band} \sim 2\ {\rm deg}$) during a single time dump (2-second, a snapshot). One `box', an area of approximately 300 sq. deg., is observed for about 2 hours (an`epoch'). The same box is observed with a rising sky and setting sky as these scans cross each other and provide us with good sky coverage in the region of overlap (Fig.~\ref{fig:scan}). This strategy is adopted to accommodate the challenging requirements for the total power \HI~IM observations \citep{Wang2021}.

\begin{figure}
    \centering
    \includegraphics[width=0.99\linewidth]{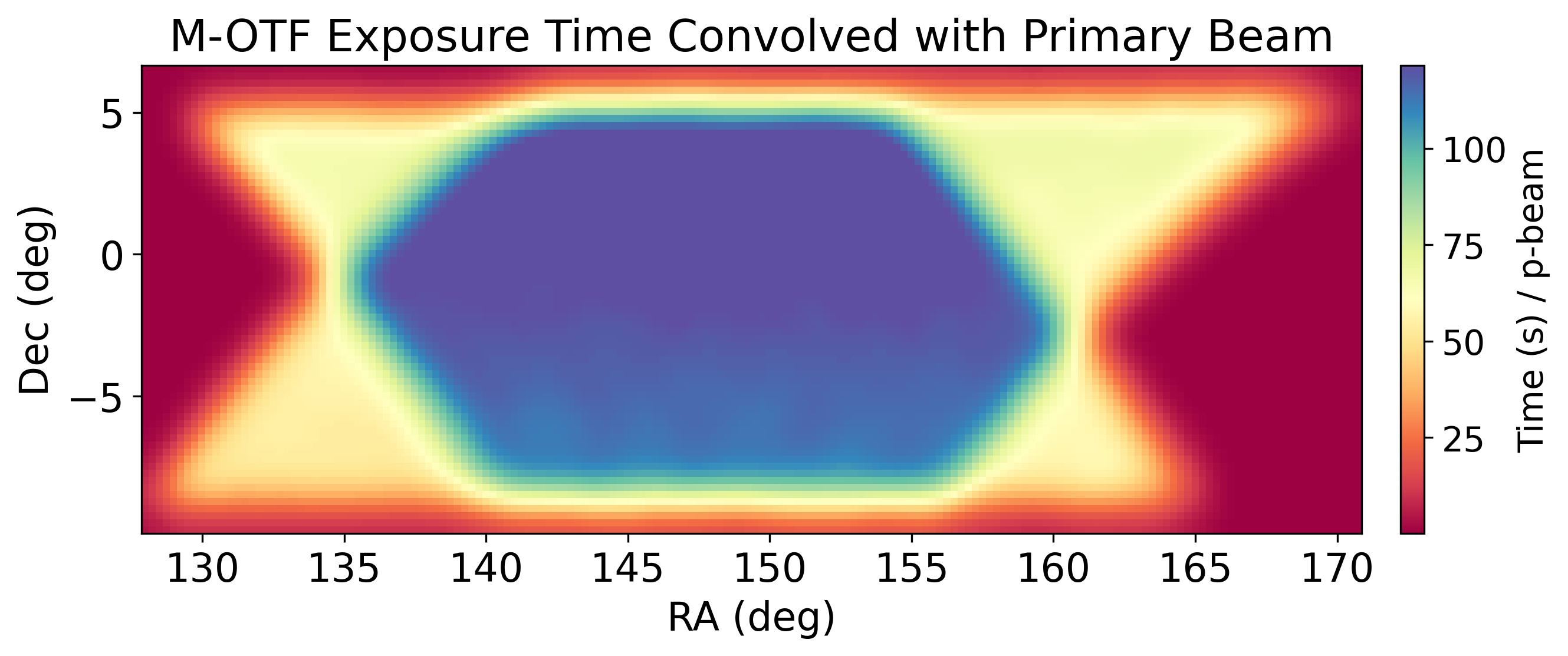}
    \caption{The figure shows the effective exposure time convolved with the primary beam, assuming every 2-second snapshot contributes to the imaging. Here we have considered a combination of one rising and one setting scan.}
    \label{fig:sen2}
\end{figure}

\begin{figure}
    \centering
    \includegraphics[width=0.99\linewidth]{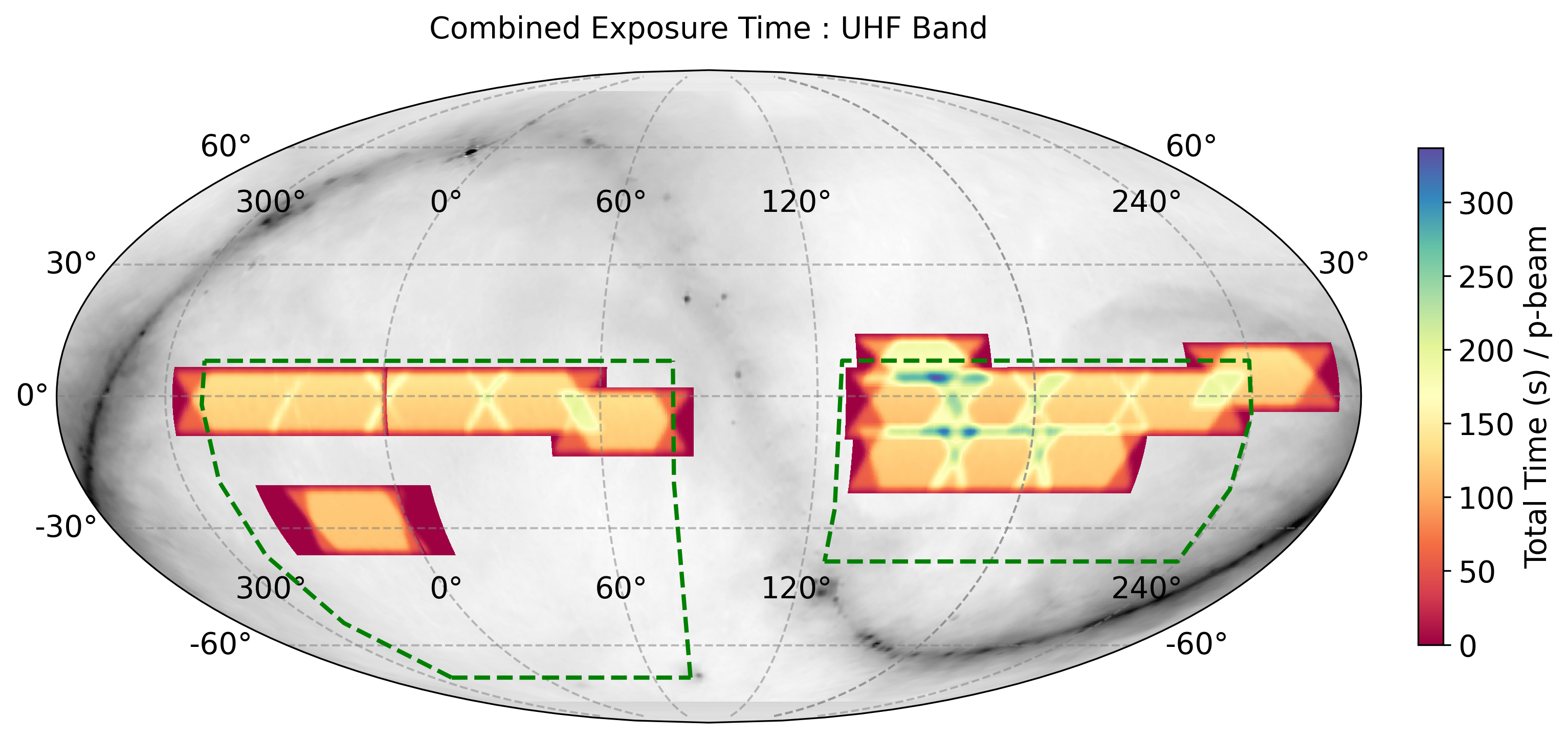}
    \caption{The green dashed lines show the planned survey region for the MeerKLASS. Background shows the Haslam map at 408 MHz \citep{Haslam1982}, we avoid observing the region close to Galactic plane. The effective exposure time convolved with the primary beam is shown for the region observed till the end of 2025. Here we have considered a combination of one rising and one setting scan to compute the effective exposuer time.}
    \label{fig:sfov}
\end{figure}

\autoref{fig:scan} shows an example scenario where one rising and one setting epochs are observed for one `box'. We perform about 26 repeated observations for each box to meet the requirements of \HI~IM observations. \autoref{fig:sen2} shows a prediction of our pipeline when combining (mosaicking) the 2-second pointings for the combination of one rising and one setting epochs from one observation box.  To estimate this we assumed a Gaussian primary beam with ${\rm FWHM} = 2^{\circ}$, which is independent of frequency. In practice primary beam varies with frequencies and contain side lobes; however, for simplicity, we ignore this here. We perform a convolution of the primary beam and the scanning pattern shown in \autoref{fig:scan} to compute this. The beam weighted exposure time gives us an optimistic estimate of the expected integration time at different RA-DEC while combining all the different 2-second measurements from an observation box. Following the radiometer equation, the expected average r.m.s. achievable in continuum ($\sigma_N$) change with the integration time by an inversely proportional relation ($\sigma_N \propto 1/\sqrt{\dt}$) and as a consequence we expect the M-OTF continuum images to be most sensitive in the hexagonal region in the middle (see \autoref{fig:scan})  where the average time spent is larger when compared with the edges of the observation box. However, as we continue with the survey the edges from different observation boxes are expected to overlap and provide us with relatively uniform r.m.s. throughout the survey area. \autoref{fig:sfov} illustrates the planned sky coverage of the ongoing MeerKLASS survey (green dashed outline). The grey-scale background map shows Galactic synchrotron emission, highlighting the low-foreground regions targeted by the survey. The foreground heat map represents the beam-weighted exposure time, demonstrating how the overlapping scanning strategy achieves uniform sensitivity across a large area of the southern sky.

In the current setup of MeerKAT, we are able to measure interferometric visibilities in two different methods. 1) The delay center of the observation can be set at a fixed RA-DEC for each scan, and during that scan, that fixed RA-DEC is used for sidereal tracking. 2) The delay center can be set to a fixed azimuth-elevation (az-el) at the beginning of the constant elevation scans and remain fixed throughout the fast scanning observation without sidereal tracking. These two setups introduce unique challenges to process and provide good quality science ready data products. We discuss how we have overcome these challenges in subsequent sections. It is worth mentioning that these setups were an unfortunate necessity due to the capabilities of MeerKAT correlators and future SKA-Mid will allow sidereal delay tracking during fast scanning observations. As a consequence, we expect these issues to be absent during SKA-Mid observations.

%The figure shows a rather optimistic scenario where we combine each 2-second snapshot measurements from a box and compute the primary beam weighted exposure time at different RA-DEC.

% Considering the visibilities, the delay center for this observation are set to a fixed az-el  at the beginning of the constant elevation scan and kept fixed throughout the observation.\KG{You should perhaps emphasise that this was an unfortunate necessity due to capabilities of the MeerKAT correlator and that SKA-MID will allow sidereal delay tracking during fast scanning observations }

\subsection{M-OTF with delay fixed in Sky coordinates}
This is a newly implemented M-OTF setup that allows to perform tracking of a RA-DEC positions that drift through the middle of the current linear scan (see \autoref{fig:scan}). This results in a separate fixed RA-DEC target for each scan line, which is tracked for the duration of the scan. This survey mode has been developed recently and is still undergoing testing.
\subsubsection{Theoretical formulation}

In a radio interferometric observation visibilities are the primary measured quantity. Considering an observation at time $t$ and frequency $\nu$ the visibility function for two antennas can be written as,
\begin{equation}
    \visbnut = \int d\Omega \, \Irnut \Brnut e^{- 2 \pi i \nubc \bt \cdot \r}.
    \label{eq:vis_def1}
\end{equation}
Here $\Irnut$ is the specific intensity of the sky at an arbitrary direction $\r$, $\Brnut$ is the primary beam response of the antennas along $\r$ and $\bt$ is the baseline vector between two antennas. For convenience, we assume a coordinate system that is fixed to the celestial sphere.   
In this case, we consider fixing the delay in the sky coordinate while the telescope is performing a fast scan. In case of M-OTF observation, let us assume that the correlator is adding delay tracking as described above towards a sky direction $\r_d$ (and that this is continuous for simplification). When averaging over time and frequency, we get:
% \begin{multline}
%     \visbnutm = \frac{1}{\dt \dnu} \int d\Omega \, \int_{\nu_m - \dnu/2}^{\nu_m + \dnu/2} \, d\nu \int_{t_m - \dt/2}^{t_m + \dt/2} dt\\
%     \times \Irnut \Brnut e^{- 2 \pi i \nubc \bt \cdot (\r-\r_d)},
%     \label{eq:vis_def2}
% \end{multline}
% 
\begin{equation}
    \visbnutm = \int d\Omega \, \Irnutm \Brnutm \times e^{- 2 \pi i \numbc \btm \cdot [\r - \r_d]} \, \sinc(\delta \Psi) \, \sinc(\delta \Phi),   
    \label{eq:vis_def4}
\end{equation}
where $t_m$ and $\nu_m$ are the central values of the intervals. The averaging in time and frequency introduces smearing where the smearing coefficients, $\delta \Psi$ and $\delta \Phi$ are expressed as \citep{Smirnov2011, Tasse2018},
\begin{align}
    &\delta \Psi  = \pi \dt \numbc \dbdtm \cdot [\r-\r_d] \label{eq:vis_def5a} \\
    &\delta \Phi = \pi \frac{\dnu}{c} \btm \cdot [\r - \r_d].
    \label{eq:vis_def5b}
\end{align}

\subsubsection{Simulated predictions}

 % It is worth mentioning that the dishes are moving back and forth at a constant elevation while recoding the visibilities. This introduces an unique set of challenges for continuum imaging that are different from traditional tracking observation. We discuss this in details at a later section.
 
To investigate the impact of the smearing and assess the accuracy of the theoretical formulation, we simulate M-OTF observations for a single source of 1~Jy at the phase-center without adding system noise. The simulations were performed around a fixed arbitrary time $T_{0}$. To capture the smearing effect, we generate visibilities with an artificially high time resolution of $\Tilde{\delta t} = \dt/100$ and frequency channel width of $\Tilde{\dnu} = \dnu/100$ for a bandwidth of 1 MHz. An unitary primary beam is assumed. We then vector sum these visibilities to the equivalent channel and time resolution to that of the actual observation. 
 \begin{figure*}
    \includegraphics[width=.99\linewidth,trim={0.0cm 0.cm 0.cm 0},clip]{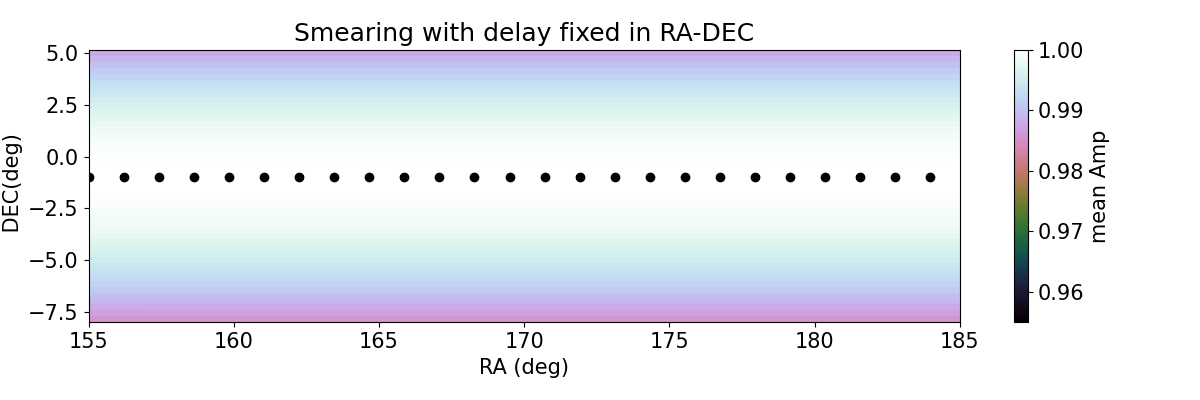}
    
    \caption{This shows the impact of the M-OTF smearing on a point source average amplitude as a function of RA-DEC when the  delay is fixed at a sky coordinate.}
    \label{fig:sm1}
\end{figure*}
\autoref{fig:sm1} shows the result, for one such block of observation, where the black points show the delay centers and the average amplitude is plotted in color-bar for a range of RA-DEC values. If there were no smearing present in the observation, we expect the average amplitude value to be unity everywhere whereas we see the minimum value of the amplitude can go upto 0.97 which implies about $3\%$ loss of flux towards the edge of the block. This can be mitigated by correcting the visibilities by the relation introduced in \autoref{eq:vis_def5b}.

\subsection{M-OTF delay fixed in Azimuth and Elevation}

\subsubsection{Theoretical formulation}

% The primary technical drawback of the current MeerKLASS observation setup is the inability to track the sky within the integration time interval ($\dt$). In the present survey mode, the delay remains fixed at an az-el during the entire constant elevation scan. During the $\dt$ interval, the sky rotates but the per-baseline correlations do not track this, leading to baseline dependent reduction of the visibility amplitudes, which is manifested as a smearing effect in the M-OTF images. As the antenna voltages have already been correlated and averaged, this effect is irreversible and cannot be fixed by correcting the delays. 

The primary drawback of the current MeerKLASS setup is its inability to track the sky during the integration time ($\dt$) due to fixed azimuth-elevation delays. As the sky rotates, this lack of tracking causes a baseline-dependent loss of phase coherence during averaging. This manifests as irreversible smearing in the M-OTF images that cannot be fixed with post-hoc delay corrections \citep{Chatterjee2025}.

The mathematical details of the smearing effect are described in \citet{Chatterjee2025}, and the smearing can be quantified by

\begin{align}
    &\delta \Psi  = \pi \dt \numbc \dbdtm \cdot \r_p  \label{eq:dpsi} \\
    &\delta \Phi = \pi \frac{\dnu}{c} \btm \cdot [\r_p - \rcm] \label{eq:dphi} , 
\end{align}
where $\btm$ is the baseline vector at time $t_m$, $\r_p$ the source position and $\rcm$ the time dependent delay position. Here $\dt = 2{\rm seconds}$ and, $\dnu = 133\, {\rm and} \, 209 {\rm kHz}$ for UHF and L-band respectively. Since the delay applied within the time resolution $\dt$ remains constant, we expect the $\sinc$ functions will reduce the amplitude of the source flux due to phase errors (smearing). The smearing can be estimated by the factor $\sinc(\delta \Psi) \sinc(\delta \Phi)$ defined in \autoref{eq:dpsi} and \ref{eq:dphi}. Using this we can construct an effective smeared PSF that can be used to create an accurate sky model from the smeared M-OTF observations (see section~\ref{subsec:img}).  

% The $\sinc$ functions will reduce the amplitude of the source flux. Naively, we could try to correct this effect by just dividing the measured visibilities by these $\sinc$ functions. But in practice, a proper deconvolution will be required as described later (see section~\ref{subsec:img}). 

\subsubsection{Simulated predictions}
 
 The key difference between a tracking and current M-OTF observation is that the delays are not corrected for Earth's rotation within the integration time and as a consequence, the aforementioned source moves away from the delay center (which ideally should be the same as the phase center). The point source is at the phase-center of the observation at the time $T_0$, and incorporating the Earth's rotation, visibilities simulated between $T_0 - \dt/2$ to $T_0 + \dt/2$ are summed to obtain the final visibility. Since the delay applied within the time resolution $\dt$ remains constant, we expect the summed visibility amplitude at large baselines to decrease from unity due to phase errors. This is in contrast to a tracking observation where we expect phases of all the component visibilities to be the same and so the amplitudes of the visibilities to be unity at all the observed baselines. 
  \begin{figure}
  \centering
    \includegraphics[width=0.99\linewidth,trim={.0cm 0.cm 0cm .0cm},clip]{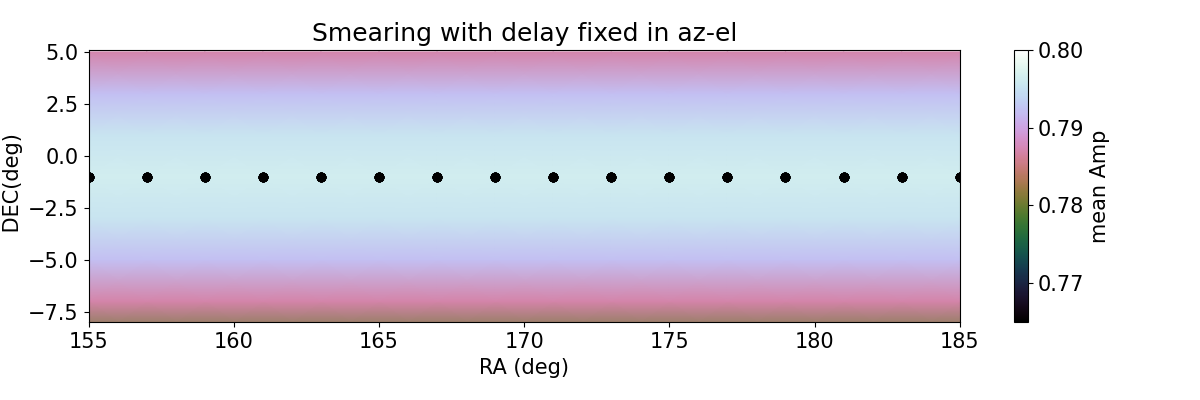}
    \caption{This shows the impact of the M-OTF smearing on a point source average amplitude as a function of steering in azimuth at a fixed elevation of $40^{\circ}$.}
    \label{fig:sm3}
% \end{figure}
  \end{figure}

During M-OTF observations, the antennas are steered $\pm 5^{\circ}$ rapidly at a constant elevation around the central azimuth. This introduces a further variation in the declination and impact of this is shown in the figure~\ref{fig:sm3}. Here the simulation is performed for a fixed elevation of $40^{\circ}$. The variation of the declination depends upon the elevation and the time of observation. Both being fixed in the simulation the effect of the smearing is primarily associated with the variation in declination due the steering of the antennas and the change in the projected baselines. These simulations give us an idea about the degree of impact on the images due to the smearing.

\section{M-OTF pipeline and imaging}\label{sec:pipe}
% \begin{figure*}
%     \includegraphics[scale=0.5]{figures/flowchart.png}
    
%     \caption{This shows the flowchart of MeerKLASS OTF end-to-end pipeline.}
%     \label{fig:flow}
% \end{figure*}
% \SCh{Some more general description of the data processing need to be added here.}
Processing of the M-OTF visibilities is different compared to the traditional tracking observation. In this section, we describe the process of flagging and calibration in our pipeline for the observed visibilities. 
\subsection{Flagging }
\subsubsection{Rogue antenna flagging}
%KG These are "rogue" antennas. Rouge antennas would be red...
The MeerKLASS scanning strategy involves driving the antennas back and forth rapidly in azimuth while the elevation remains fixed. However, we have found that during these fast scanning some of the antennas can fall behind or just stop moving for some time. It is important that we identify these misbehaving or `rogue' antennas and flag them from the visibilities. To identify these rogue antennas we compare the pointing of each antenna as a function of time with the median of the antenna pointing distribution. If an antenna deviates more than $\delta \theta \sim 0.1^{\circ} $ from the median pointing of the antennas at any time stamp, it is flagged. 

\subsubsection{RFI flagging}
Radio Frequency Interference (RFI) is emitted from various sources, terrestrial or orbital, and creates a nuisance for ground-based radio observations.  Since MeerKAT has a large effective collecting area of approximately 1960.88 square meters, it has high sensitivity and as the sensitivity of the instruments increases, so does the sensitivity to unwanted signals \citep{Harper2018, Engelbrecht2024}. Primary RFI contributors to the MeerKLASS observations are Digital TV (UHF), GSM (Mobile phones) (UHF + L-band), Aircraft transponders, GPS, GLONASS, Galileo and Inmarsat\footnote{\href{https://skaafrica.atlassian.net/wiki/x/AQAzEg}{https://skaafrica.atlassian.net/wiki/x/AQAzEg}}.  MeerKAT has RFI mitigation systems to prevent RFI before and during the observation \citep{Jonas2016}. However, as even the best RFI mitigation methods cannot completely prevent all RFI \citep{Baan2010}, we must employ methods to reduce the effect of RFI after observation.

After flagging the rogue antennas, we process the calibrator and the scan data through our RFI flagging step. The RFI flagging is performed using the \texttt{TRICOLOR} package \citep{Hugo2022}, which implements the `sumthreshold' algorithm \citep{Offringa2010} optimised for MeerKAT.  Considering the UHF-band, we see that the frequency window between 580 - 880 MHz is relatively clean with the flag fraction mostly below 0.15. We found the pipeline is able to remove all the major RFI contamination successfully; however, the RFI environment somewhat changes in time, and as a result, the flag fraction can change between one epoch to another for the same observation box.

\subsection{Calibration}\label{subsec:cal}
The current MeerKLASS calibration strategy relies upon a primary, a secondary and a polarisation calibrator. It is not efficient to stop and steer the telescope for calibration observation once the constant elevation scanning starts. Due to this limitation, we observe the calibrators at the beginning and toward the end of an epoch for a given observation box. We use \texttt{CARACal}\footnote{\href{https://caracal.readthedocs.io/en/latest/index.html}{https://caracal.readthedocs.io/}} \citep{caracal} to produce calibration tables from the primary, secondary and polarisation calibrators. We follow a `standard' calibration strategy to solve the delays, band-passes and gains, and calibrate for the absolute flux using the primary calibrator source. Most of the MeerKLASS observations are preformed at night and are significantly distant from the solar activity. Hence, we expect the ionospheric effects to be subdominant and these have been ignored throughout this paper.

\subsection{Phase Center Correction}
The delays applied in the correlator are calculated for a particular RA-DEC or a single az-el, whereas the antennas point in a direction which is different from the delay center. This step in our pipeline splits the whole measurement set (MS) file of an observation epoch into multiple 2s snapshot files, each of which contains visibilities taken at a single, unique timestamp. Further, it applies a phase-rotation to the visibilities from the observed delay center to the pointing center for every visibility integration time ($\delta t$). We use the \texttt{CHGCENTER} task in \texttt{WSClean}\footnote{\href{https://wsclean.readthedocs.io/en/latest/index.html}{https://wsclean.readthedocs.io/}}.

Finally, we use ``on-the-fly" calibration option in \texttt{CARACal} to apply the calibration tables obtained previously (Section~\ref{subsec:cal}) on each 2s delay rotated snapshot measurement sets.  At the end of this step, the calibrated snapshot visibilities are stored on disc and further processed for imaging. 

\subsection{Imaging}\label{subsec:img}
\begin{figure*}
\hspace{-0.75cm}
  \begin{subfigure}{.99\textwidth}
  % \centering
    \includegraphics[width=.99\linewidth,trim={0.cm 0.cm 0.cm 0},clip]{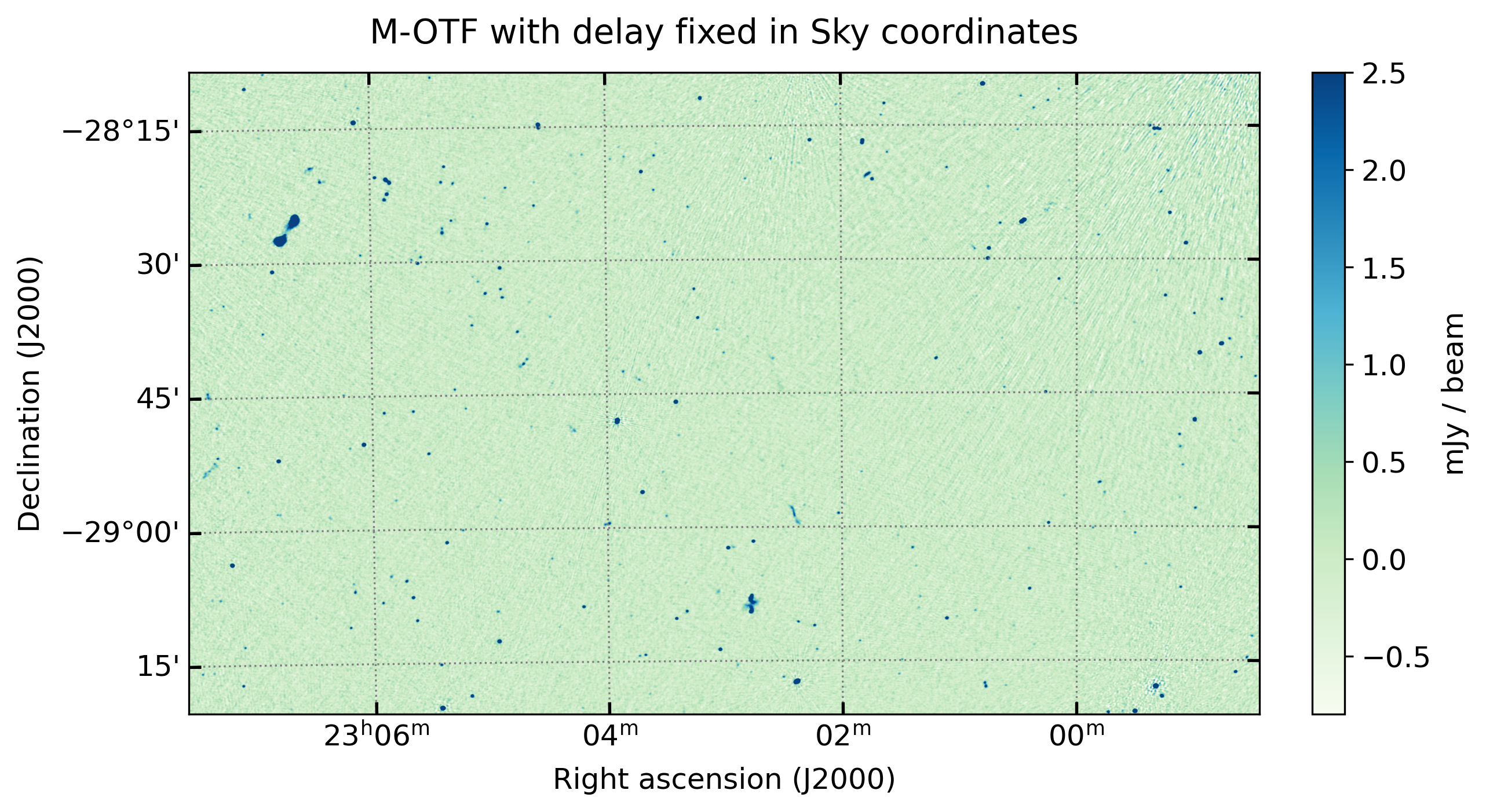}
    % \caption{First subfigure}
  \end{subfigure}%
  % \hspace{-1cm}
  \\
  
  \begin{subfigure}{.99\textwidth}
  % \centering
  \hspace{-0.75cm}
    \includegraphics[width=.99\linewidth,trim={0.0cm 0.cm 0.0cm 0.cm},clip]{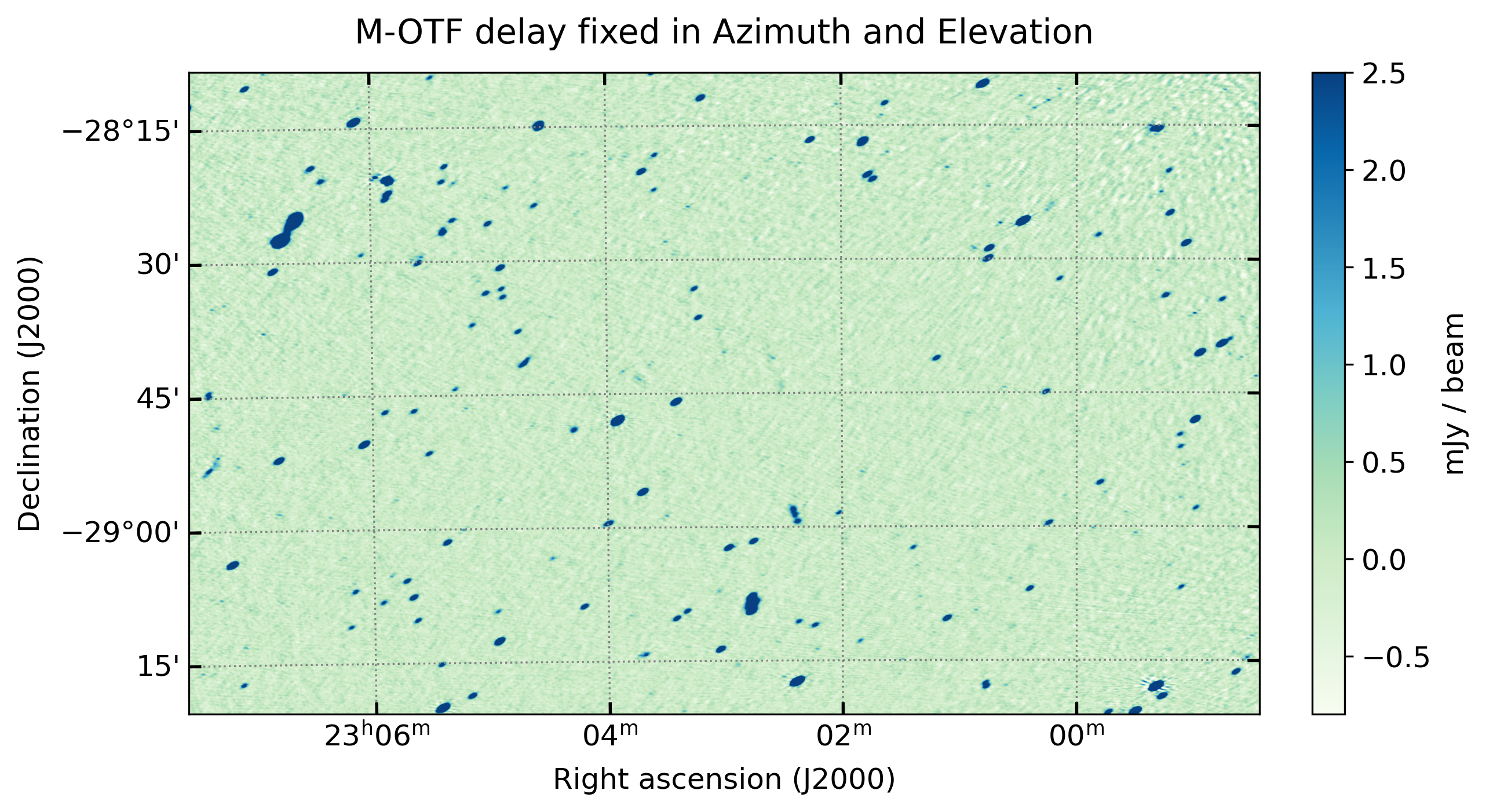}
    % \caption{Second subfigure}
  \end{subfigure}
    % \bigskip
    \caption{These show the M-OTF final image products in the UHF band. Top panel show the image quality for the newly implemented setup where the delay for each scan is set in the sky coordinate and the correlator performs sidereal tracking to accommodate Earth's rotation. The bottom panel shows the image quality for the setup where the delay is fixed at an az-el and the telescope doesn't tracks the Earth's rotation.}
    \label{fig:mos1}
\end{figure*}
The immediate and most intuitive product for M-OTF are the 2s total intensity (Stokes I) snapshot images. The 64 dishes of the MeerKAT, in conjunction with a cryogenically cooled receiver system, provide excellent snapshot $u$-$v$ coverage and low system temperature ($T_{\rm sys}$), which are essential for imaging the 2s snapshots. To achieve better depth sensitivity and image fidelity, we have chosen to perform visibility plane mosaicking (joint deconvolution) for the M-OTF images (see \citealt{Chatterjee2025}). We use \texttt{DDFacet} \citep{Tasse2018} to perform smearing correction and imaging the M-OTF visibilities with mosaicking in the visibility plane. The purpose of using an algorithm that relies on faceting is to approximate a wide FoV with many narrow field images.

In this chapter, we show the image products using both the MeerKAT set ups. The top panel of the \autoref{fig:mos1} shows the image from the newly implemented setup for M-OTF with delay fixed at sky coordinate with sidereal tracking, whereas the bottom panel shows the resulting image from M-OTF delay fixed at an az-el without sidereal tracking. 
Considering the top panel, we have used observations from one epoch only, where  we have combined 31 snapshots per square degrees to construct a $3.2^{\circ} \times 3.2^{\circ}$ image and here we show a central cutout of $1^{\circ} \times 1^{\circ}$. Using the visibility plane mosaic,we are able to achieve a continuum resolution of $14''$ and a noise of $\sim 163 \mu {\rm Jy/beam}$ in the UHF-band.

The bottom panel of \autoref{fig:mos1} shows a visibility plane mosaic image from the M-OTF observation where the delay is fixed in $az$-$el$ and the correlator does not track Earth's rotation. We used the DDFacet to correct for the smearing during the deconvolution with a smeared PSF. We are able to achieve a continuum resolution of $32'' \times 15''$ and a noise of $\sim 160 \mu {\rm Jy/beam}$ in the UHF-band. The resolution is significantly poorer when compared with newly implemented observing mode due to smearing.

\section{UHF-band results} \label{sec:uhfband}

\begin{figure*}
\centering
\includegraphics[width=\textwidth]{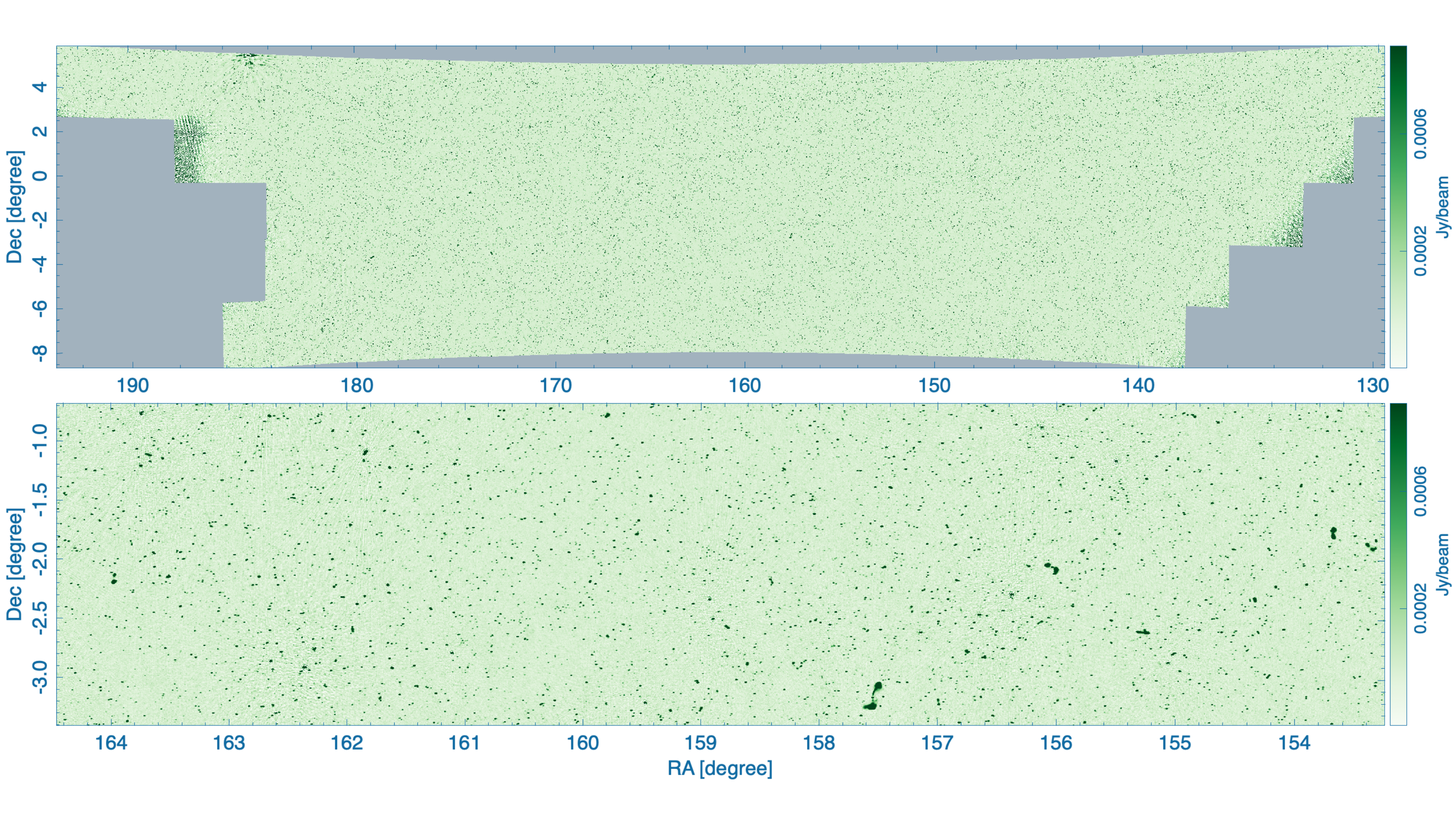}
\caption{
Overview of the mosaic and image quality across the MeerKLASS UHF survey area used in DR1.
The top panel shows the full $\sim$800\,deg$^2$ continuum mosaic constructed from 89 tiles at the UHF band centre (816 MHz). Bottom panel displays a zoomed in view of the survey's center region.
}
\label{fig:uhf_mosaic}
\end{figure*}

The MeerKLASS OTF UHF survey constitutes the largest wide-area interferometric imaging effort using fast-scanning data from MeerKAT’s 544–1088\,MHz band. It provides a unique window on the faint radio sky, bridging classical L-band extragalactic surveys and the low-frequency regime probed by instruments such as LOFAR and MWA. The overarching objective of the MeerKLASS UHF survey is to deliver a deep, wide-area, high-resolution radio continuum map of the southern sky while running fully commensally with the single-dish H\,\textsc{i} intensity-mapping experiment.

The first data release (DR1, \citealt{Paul2025}) focuses on a subset of approximately 800\,deg$^{2}$ obtained from eight OTF observing blocks (corresponding to $\sim$12\,h of total integration). The footprint overlaps the DESI extragalactic field and serves as a pilot region for future survey expansion. We divided the survey area into 89 individual tiles for imaging (each $\approx 3^\circ\times3^\circ$). Direction-independent imaging and self-calibration were performed with \textsc{DDFacet} using robust $0$ weighting. The resulting continuum mosaic (Fig.~\ref{fig:uhf_mosaic}) reaches a sensitivity of $\approx 35~\mu$Jy\,beam$^{-1}$ in regions with contribution from 8 datablocks and a median synthesized beam of $32''\times17''$ (FWHM). The achieved depth is consistent with thermal-noise expectations for the cumulative integration time, demonstrating that stable, high-fidelity imaging is attainable despite the intrinsic motion of the OTF scanning strategy.

\begin{figure}
    \centering
    \includegraphics[width=1\linewidth]{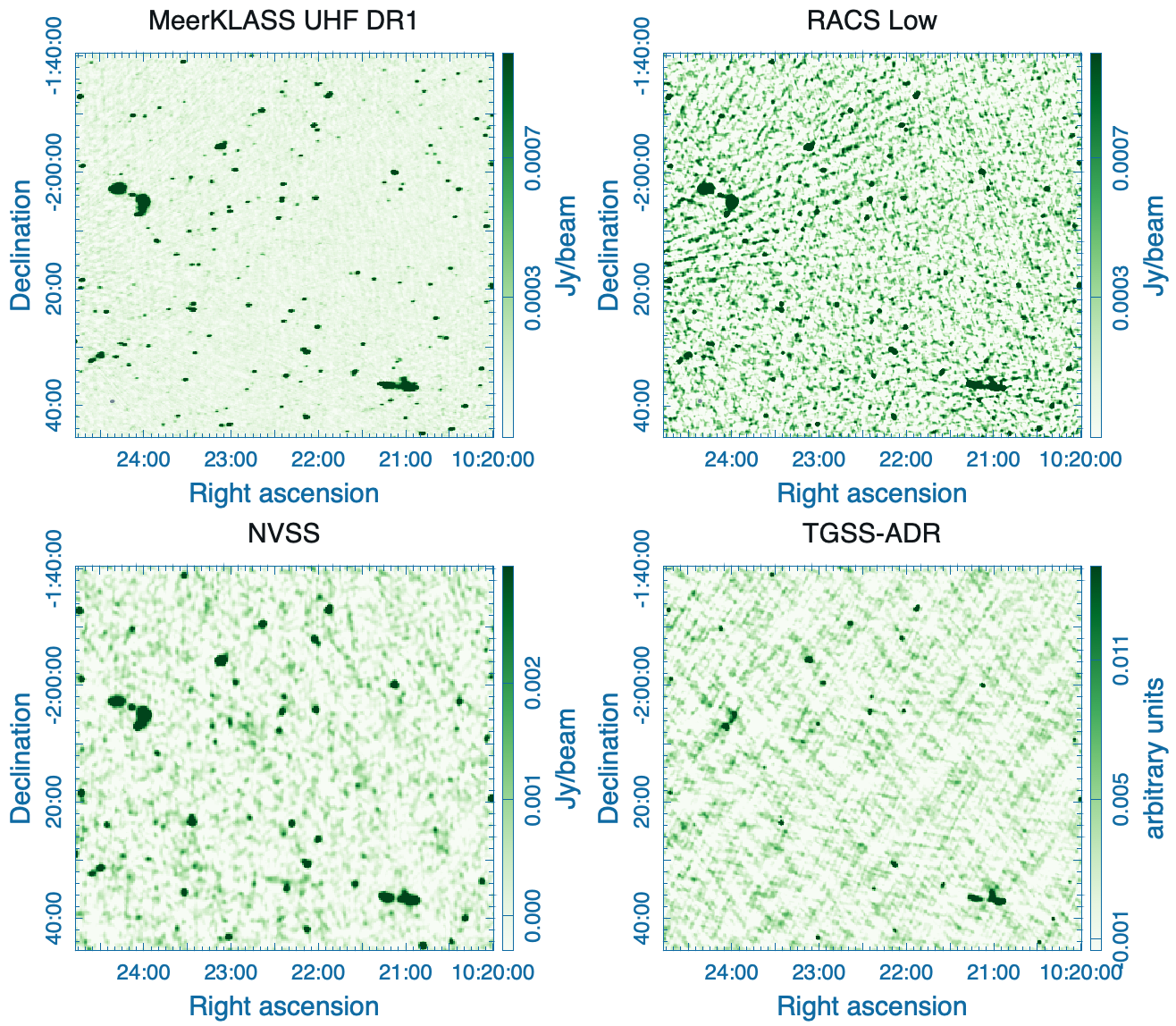}
    \caption{%UHF-band image comparison with other surveys; 
    Comparison of the MeerKLASS UHF-band image with major radio surveys.  Flux-density scales differ between panels and are set by each survey’s sensitivity (see \autoref{tab:survey_SH}). The MeerKLASS UHF-band DR1 image reveals substantially fainter sources than the other surveys.}
    \label{fig:UHFband_racs}
\end{figure}

\autoref{fig:uhf_mosaic} provides a global view of the continuum image quality across the entire 800 deg$^2$ footprint. The top panel displays a composite mosaic constructed from the restored images of 89 individual tiles. This full-field image was assembled using \textsc{Montage}\footnote{\url{http://montage.ipac.caltech.edu}}, which reprojects each input FITS image to a common WCS frame before co-adding them into a seamless large-area mosaic. To produce this, we first collected the final restored images for all tiles, each corresponding to a $3.2^\circ \times 3.2^\circ$ field generated by a single DDFacet imaging run. All tiles share a common pixel scale of 3 arcsec and are aligned in equatorial coordinates. The bottom panel shows a zoomed in view of the survey center. \autoref{fig:UHFband_racs} shows a comparison of M-OTF image with RACS-Low \citep{Hale2021}, NVSS \citep{Condon1998} and TGSS-ADR \citep{Intema2016}. These demonstrate the high image quality achieved across the survey, revealing numerous compact and extended sources, clean background regions, and minimal residual artefacts.
Source extraction with \textsc{PyBDSF} yields a catalog of $\sim$\,10$^{5}$ sources at a $7\sigma$ threshold. Morphological statistics indicate that the population is largely marginally resolved. The total-to-peak flux-density ratio ($T/P$) declines from $\sim 1.12$ at $\mathrm{SNR}\sim10$ to $\sim 1.03$ by $\mathrm{SNR}\sim 10^{3}$, and the resolved fraction similarly decreases with SNR, consistent with a catalog dominated by compact sources and minimal resolution bias at high SNR.

Differential source counts derived from DR1 are in good agreement with existing wide surveys (e.g.\ RACS-Low, NVSS) after frequency scaling to a representative UHF band centre (816 MHz), supporting both the photometric fidelity and the completeness of the catalog above $\sim 1$\,mJy. Astrometric offsets are well below the $3''$ pixel scale, and fluxes agree with external surveys within expected spectral and resolution systematics. Collectively, these results validate OTF interferometric imaging as a robust and efficient mode for wide-field radio continuum surveys. 

\section{L-band results} \label{sec:lband}
%Results from MeerKLASS DR1 (Sarvesh)

\begin{figure}
    \centering
    \includegraphics[width=.9\linewidth,trim={0.cm 0.cm 0.cm 0},clip]{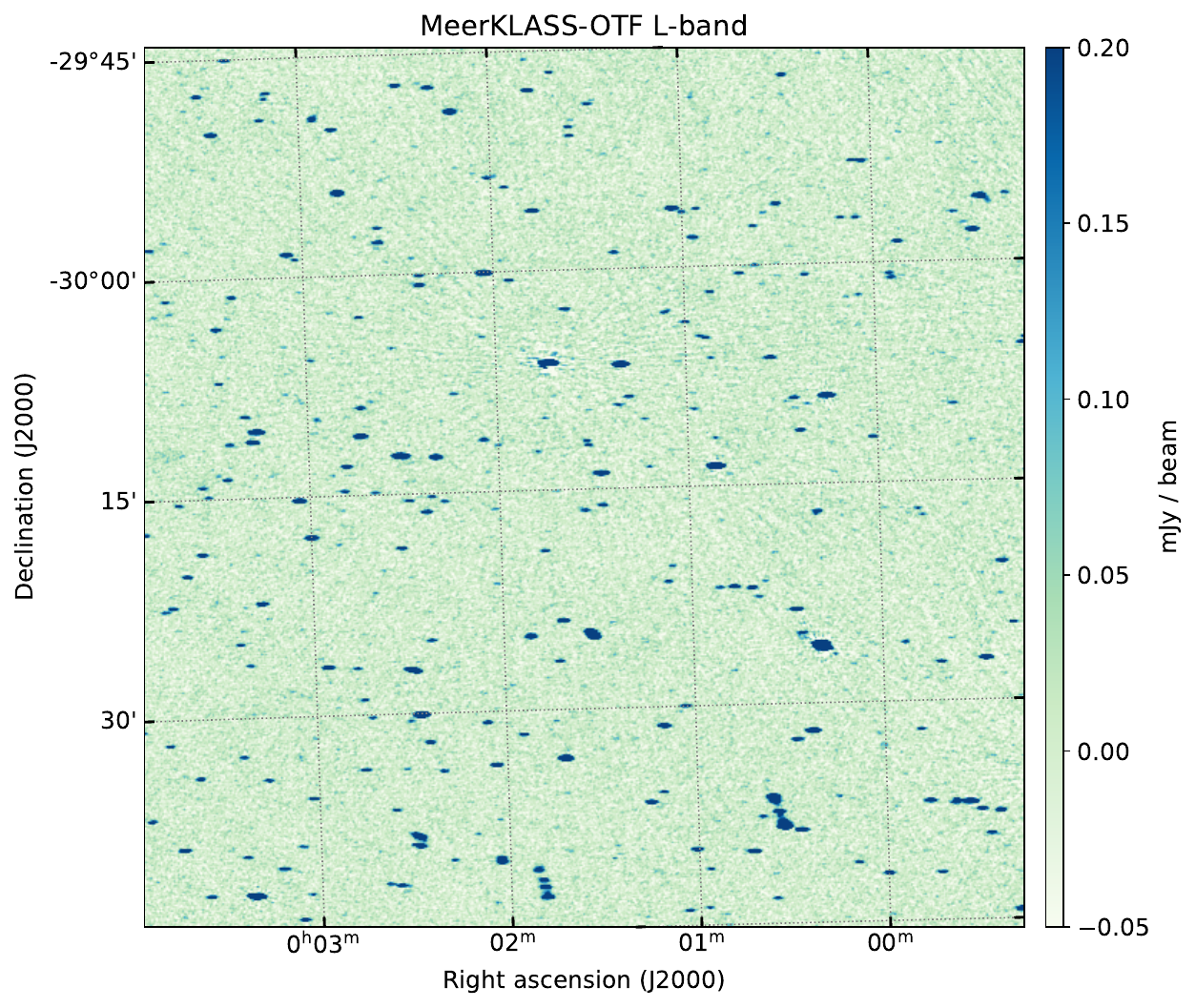}
    \caption{MeerKLASS L-band image at 1.284\,GHz, covering a one square degree field. The map reaches an RMS noise of 32.6\,$\mu$Jy/beam. The restoring beam is $26''\times 7.8''$, giving the elongated resolution pattern you see in the map.}
    \label{fig:Lband_1deg}
\end{figure}

The MeerKLASS L-band survey employs the fast-scanning OTF observations to cover approximately
$\sim$270 deg$^{2}$ of sky across the $856-1712\,$MHz band. This initial data release (DR1, \citealt{Mangla2025}) is compiled from 67 individually processed image tiles, with each tile covering an area of $\approx 2^\circ\times2^\circ$ on the sky. The survey's footprint provides significant overlap with existing wide-area optical surveys, such as Kilo-Degree Survey (KiDS-DR5, \cite{KiDS_DR5}), providing a valuable multi-wavelength dataset. %https://ui.adsabs.harvard.edu/abs/2024A&A...686A.170W
%and the upcoming DESI Legacy DR11 \SM{add the reference} don't have the link. 

Following the process described in \autoref{sec:pipe}, the resulting map has an excellent, consistent quality, achieving an average rms noise level of $\approx\,34\,\mu\rm Jy$ across the entire field. The median resolution of a synthesized beam is $26''\times8''$. The achieved imaging depth aligns precisely with the expectations from the thermal noise limit for the cumulative observing time, thus validating the efficiency and robustness of the OTF observing strategy. A representative section of the resulting continuum image, which highlights the high quality and low noise floor achieved, is shown in \autoref{fig:Lband_1deg}. 

\begin{figure}
    \centering
    \includegraphics[width=1\linewidth]{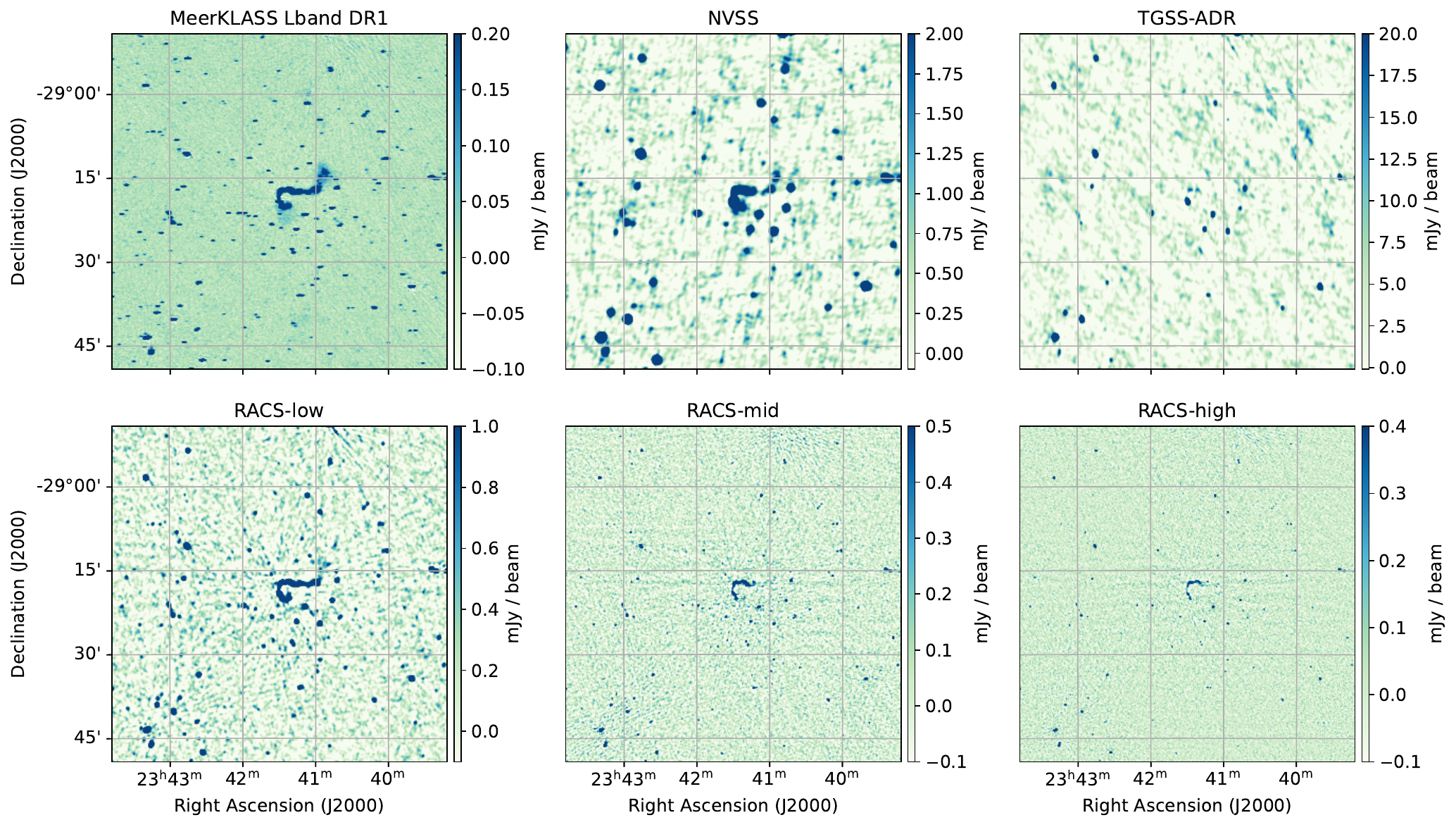}
    \caption{Comparison of the MeerKLASS L-band image with major radio surveys. Top panel: MeerKLASS L-band DR1 image with NVSS and TGSS-ADR overlaid for the same sky region. Lower panels: RACS-low, RACS-mid, and RACS-high images of the same field. Flux-density scales differ between panels and are set by each survey’s sensitivity (see \autoref{tab:survey_SH}). The MeerKLASS L-band DR1 image reveals substantially fainter sources and more extended emission than the other surveys.
    %The MeerKLASS L-band DR1 image (RMS sensitivity $\sim 33\,\mu\mathrm{Jy},\mathrm{beam}^{-1}$) is $\gtrsim 14\times$ deeper than NVSS ($\sim 450\,\mu\mathrm{Jy}\mathrm{beam}^{-1}$) and $\sim 4$–$9\times$ deeper than the RACS-low/mid/high images ($\sim 0.14$–$0.26\,\mathrm{mJy}\mathrm{beam}^{-1}$), revealing substantially fainter sources and extended emission.
    }
    \label{fig:lband_racs}
\end{figure}

Source detections was performed using the \textsc{PyBDSF} source finding algorithm, yielding over 40,000 sources detected above $\mathrm{S/N}$ threshold of 9. This catalog is estimated to be over 90\% complete above a flux density of 0.6\, mJy. To confirm the photometric and astrometric fidelity of the source measurements, the properties of the detected sources were cross-matched against established external radio surveys, including the NVSS, TGSS-ADR, and the ASKAP-RACS catalogs (Low, Mid, and High). Comparison with the RACS-mid is illustrated in \autoref{fig:lband_racs}, confirming the consistency of the MeerKLASS flux scale. We found that the astrometric positions are highly accurate, with positional errors being less than 1.5$''$ (the map's pixel size). Furthermore, the flux density measurements agree with those from external surveys to within a 2-3 per\,cent uncertainty.  This agreement is based on a global fit to the results from all common detected sources between the surveys, after applying frequency corrections based on a typical synchrotron spectrum ($S_{\nu} \propto \nu^{\alpha}$ with $\alpha = -0.7$). Collectively, these initial results validate the OTF imaging technique as a dependable and efficient method for executing wide-scale radio surveys with a successfully achieved low noise floor.

\section{Predictions for SKA-Mid} \label{sec:ska}

\begin{table}[]
\centering
\begin{tabular}{|l|c|c|c|c|}
\hline

Configuration                                                                             & MeerKAT            & \begin{tabular}[c]{@{}l@{}}SKA-Mid \\ (64 MeerKAT)\end{tabular} & \begin{tabular}[c]{@{}l@{}}SKA-Mid\\ (AA*)\end{tabular} & \begin{tabular}[c]{@{}l@{}}SKA-Mid\\ (AA4)\end{tabular} \\ \hline \hline
Dishes                                                                                    & 58                 & 64                                                              & 80+64                                                   & 133+64                                                  \\ \hline
Synthesize Beam                                                                           & $14.4'' $ & $15.89''\times 10.69''$                                             & $3.49'' \times 2.94''$                                      & $1.32'' \times 1.14''$                                      \\ \hline
\begin{tabular}[c]{@{}c@{}}Stokes I snapshot noise \\ (2-second) in $\mu{\rm Jy}/{\rm beam}$ \end{tabular}             & 277               & 343                                                             & 142.6                                                   & 100                                                     \\ \hline
\begin{tabular}[c]{@{}c@{}c@{}}Stokes I one epoch noise\\ (18-seconds / sq. deg) \\ in $\mu{\rm Jy}/{\rm beam}$ \end{tabular} & 104              & 114                                                              & 48                                                      & 33                                                      \\ \hline
\begin{tabular}[c]{@{}c@{}}Stokes I 26 epoch noise  \\ in $\mu{\rm Jy}/{\rm beam}$ \end{tabular}                                                                 & 35                 & 30                                                              & 9                                                       & 7                                                       \\ \hline
\end{tabular}
\caption{The table shows forecast for OTF mapping with SKA-Mid.}
\label{tab:2}
\end{table}

In this section we discuss our forecast from SKA-Mid OTF observations. Considering SKA-Mid is a general purpose telescope, a good all-sky radio source model is an absolute necessity. The OTF mapping technique developed using SKA-Mid pathfinder telescope MeerKAT has a high survey speed, able to achieve uniform sensitivity over a large sky area and can be performed in a commensal observation with \HI~IM experiments. This makes the OTF mapping suitable for all-sky model building and long period transient search.

For the forecast, we have assumed SKA-Mid telescope will be deployed in three stages. In the first stage, all the existing 64 MeerKAT dishes will be updated to SKA-Mid specifications. Second stage (SKA-Mid AA*) will be a combination of 80 dishes with 15\,m diameter and 64 dishes with 13.5\,m diameter. Finally, the third stage (SKA-Mid AA4) will contain 133 dishes of 15\,m diameter along with the 64 MeerKAT dishes. The new 15\,m dishes will help SKA-Mid to reach high angular resolution while remaining sensitive for large-scale emission due to the existing dens coverage of baseline $\leq 1$Km. 

\autoref{tab:2} shows the forecast for OTF mapping with SKA-Mid. For completeness, we have also included existing MeerKAT predictions to the table also. From our current OTF setup, we combine about 9  snapshots per square degree per epoch to perform visibility plane mosaicking, and we have used this for the forecast. For uniformity, we have assumed that 70\% of the observed data is usable. It is worth noting that our current M-OTF observations are performed at night only, and we find that about 15\% data from the fast-scan is being flagged. This number can vary day-to-day basis, but rms estimates in the table shown here are poorer than what we actually see. Further following our current M-OTF plan, we assume that we will be combining 26 epochs per box with SKA-Mid. Using this rather pessimistic scenario, we expect to achieve an rms of 9 $\mu{\rm Jy}/{\rm beam}$ with a resolution of $3.2''$ using AA* and  $7 \mu{\rm Jy}/{\rm beam}$ with a resolution of $1.2''$ using AA4 (\autoref{fig:comp_sensitivity}). Further, we expect the SKA-Mid setup will be able to update the delay center at an interval of 0.5-seconds, making the OTF mapping practically smearing free. However, combining two different antenna design may pose some challenge toward imaging and flux-scale accuracy. Investigates are going on to asses the impact of this and will be addressed in future studies.
\section{Summary}\label{sec:summary}
MeerKLASS is an \HI~ intensity mapping survey that aims to detect the 21cm power spectrum from redshift $<1.44$. Alongside the auto-correlation (single dish) visibilities, MeerKLASS also records interferometric visibilities in On-The-Fly (OTF) mode. In this work, we introduce the MeerKLASS OTF (M-OTF) survey. We have developed and tested a new calibration and imaging capability for MeerKAT in which antennas are moved back-and-forth at a constant elevation and the visibilities are recorded continuously at 2s interval. This commensal continuum observation significantly reduces the slew-and-settle overhead compared to a traditional tracking observation. As a consequence we achieve a high survey speed of $15\,{\rm deg}^2 {\rm hr}^{-1}$ to achieve a noise of $25\,\mu{\rm Jy beam}^{-1}$ in the UHF band. 

We have developed an end-to-end pipeline that flags, calibrates, performs OTF correction and stores the visibilities in 2s snapshot measurement sets. The relevant steps of this process are described in detail in \autoref{sec:motf}. These 2s snapshot measurement sets are independent, and standard imaging techniques can be used to produce 2s snapshot images. These are crucial for slow transient search. However, there are limitations to M-OTF observations. In the current format, the delay center of the observation is fixed during the scanning process, whereas the telescope pointing performs a horizontal raster scan, and the earth rotates. This introduces a smearing that affects our images significantly (\autoref{sec:pipe}). To overcome these limitations, we have incorporated a smearing correction during the imaging, where the dirty image is deconvolved using a smeared PSF that constructs a correct sky model. Due to this correction, the resolution of the images go down (\autoref{subsec:img}). 

In principle, the 2s images can be combined to achieve a deeper image that is required for producing a source catalog and other scientific objectives. But that limits our ability to deconvolve faint sources. To overcome this, we have used \texttt{DDFacet} to perform joint deconvolution (visibility plane mosaicking) in our imaging pipeline. Further, we have performed a thorough investigation of the source flux, positional accuracy by comparing with other overlapping available catalogs. We compare the astrometry and flux-density measurements of MeerKLASS catalog with RACS-Mid. This comparison shows that astrometric offsets are typically constrained to within a pixel of the M-OTF data. Comparison of source  flux densities show agreement with RACS-Mid observations at high SNR, whereas the scatter in the flux-density comparison increases at low SNR regime while maintaining the average one-to-one trend. 
 
MeerKLASS OTF survey is an ideal example of a commensal survey where the telescope time is utilized for multiple science drivers. This potentially be helpful for producing large area sky models and slow transient searches with the upcoming SKA-Mid. In \autoref{sec:ska} we have discussed the expected sensitivity and resolution achievable with different proposed stages of  SKA-Mid. We find that with SKA-Mid AA4 configuration, we can achieve an rms of $7\,\mu{\rm Jy beam}^{-1}$ at a resolution of $1.2^{''}$. However, it is difficult to estimate the impact of combining visibility measurement from two different type of antennas and more tests are going on to asses this.  

% we should add arxiv of all three DR1 papers. Just putting it here so I don't forget.

\bibliographystyle{abbrvnat-maxbibnames4.bst}
\bibliography{mylist} % if your bibtex file is called example.bib%

@article{Alonso2021,
    author = {Alonso, David and Bellini, Emilio and Hale, Catherine and Jarvis, Matt J and Schwarz, Dominik J},
    title = {Cross-correlating radio continuum surveys and CMB lensing: constraining redshift distributions, galaxy bias, and cosmology},
    journal = {Monthly Notices of the Royal Astronomical Society},
    volume = {502},
    number = {1},
    pages = {876-887},
    year = {2021},
    month = {01},
    issn = {0035-8711},
    doi = {10.1093/mnras/stab046},
    url = {https://doi.org/10.1093/mnras/stab046},
    eprint = {https://academic.oup.com/mnras/article-pdf/502/1/876/38869980/stab046.pdf},
}

@article{Baan2010,
  author = "Baan, Willem",
  title = "{RFI Mitigation in Radio Astronomy (invited)}",
  doi = "10.22323/1.107.0001",
  journal = "PoS",
  year = 2010,
  volume = "RFI2010",
  pages = "001"
}

@misc{Braun2019,
      title={Anticipated Performance of the Square Kilometre Array -- Phase 1 (SKA1)}, 
      author={Robert Braun and Anna Bonaldi and Tyler Bourke and Evan Keane and Jeff Wagg},
      year={2019},
      eprint={1912.12699},
      archivePrefix={arXiv},
      primaryClass={astro-ph.IM},
      url={https://arxiv.org/abs/1912.12699}, 
}

@INPROCEEDINGS{caracal,
       author = {{J{\'o}zsa}, G.~I.~G. and {White}, S.~V. and {Thorat}, K. and {Smirnov}, O.~M. and {Serra}, P. and {Ramatsoku}, M. and {Ramaila}, A.~J.~T. and {Perkins}, S.~J. and {Maccagni}, F.~M. and {Makhathini}, S. and {Moln{\'a}r}, D.~C. and {Kamphuis}, P. and {Kleiner}, D. and {Hugo}, B.~V. and {de Blok}, W.~J.~G. and {Andati}, L.~A.~L.},
        title = "{MeerKATHI - an End-to-End Data Reduction Pipeline for MeerKAT and Other Radio Telescopes}",
    booktitle = {Astronomical Data Analysis Software and Systems XXIX},
         year = 2020,
       editor = {{Pizzo}, R. and {Deul}, E.~R. and {Mol}, J.~D. and {de Plaa}, J. and {Verkouter}, H.},
       series = {Astronomical Society of the Pacific Conference Series},
       volume = {527},
        month = jan,
        pages = {635},
          doi = {10.48550/arXiv.2006.02955},
archivePrefix = {arXiv},
       eprint = {2006.02955},
 primaryClass = {astro-ph.IM},
       adsurl = {https://ui.adsabs.harvard.edu/abs/2020ASPC..527..635J}
}

@article{Chatterjee2025,
    author = {{Chatterjee}, Suman and {Santos}, Mario G. and {Rozgonyi}, Kristof and {Grainge}, Keith and {Mangla}, Sarvesh and {Mohr}, Joseph J. and {Paul}, Sourabh and {Perrott}, Yvette and {Smirnov}, Oleg M. and {Tasse}, Cyril and {Wolz}, Laura},
        title = "{The MeerKLASS On-the-Fly continuum survey: pipeline design and validation}",
      journal = {arXiv e-prints},
         year = 2025,
        month = dec,
          eid = {arXiv:2512.11978},
        pages = {arXiv:2512.11978},
          doi = {10.48550/arXiv.2512.11978},
archivePrefix = {arXiv},
       eprint = {2512.11978},
 primaryClass = {astro-ph.IM},
       adsurl = {https://ui.adsabs.harvard.edu/abs/2025arXiv251211978C}
}

@misc{Paul2025,
  title={The MeerKLASS UHF On-the-Fly Continuum Survey -- Data Release I}, 
  author={Sourabh Paul and Keith Grainge and Mario G. Santos and Suman Chatterjee and Sarvesh Mangla and Laura Wolz and Joseph J. Mohr and Oleg Smirnov and Cyril Tasse and Kristof Rozgonyi and Matthias Hoeft and Yvette Perrott},
  year={2025},
  eprint={2512.11964},
  archivePrefix={arXiv},
  primaryClass={astro-ph.GA},
  url={https://arxiv.org/abs/2512.11964}, 
}

@article{Mangla2025,
    author = {{Mangla}, Sarvesh and {Mohr}, Joseph J. and {Rozgonyi}, Kristof and {Chatterjee}, Suman and {Grainge}, Keith and {Paul}, Sourabh and {Santos}, Mario G. and {Perrott}, Yvette and {Smirnov}, Oleg M. and {Tasse}, Cyril and {Wolz}, Laura},
        title = "{The MeerKLASS L-band On-the-Fly Continuum Survey: Data Release 1}",
      journal = {arXiv e-prints},
         year = 2025,
        month = dec,
          eid = {arXiv:2512.17685},
        pages = {arXiv:2512.17685},
          doi = {10.48550/arXiv.2512.17685},
archivePrefix = {arXiv},
       eprint = {2512.17685},
 primaryClass = {astro-ph.IM},
       adsurl = {https://ui.adsabs.harvard.edu/abs/2025arXiv251217685M}
}

@ARTICLE{Condon1998,
       author = {{Condon}, J.~J. and {Cotton}, W.~D. and {Greisen}, E.~W. and {Yin}, Q.~F. and {Perley}, R.~A. and {Taylor}, G.~B. and {Broderick}, J.~J.},
        title = "{The NRAO VLA Sky Survey}",
      journal = {\aj},
         year = 1998,
        month = may,
       volume = {115},
       number = {5},
        pages = {1693-1716},
          doi = {10.1086/300337},
       adsurl = {https://ui.adsabs.harvard.edu/abs/1998AJ....115.1693C}
}

@article{Cunnington2022,
    author = {Cunnington, Steven and Li, Yichao and Santos, Mario G and Wang, Jingying and Carucci, Isabella P and Irfan, Melis O and Pourtsidou, Alkistis and Spinelli, Marta and Wolz, Laura and Soares, Paula S and Blake, Chris and Bull, Philip and Engelbrecht, Brandon and Fonseca, José and Grainge, Keith and Ma, Yin-Zhe},
    title = "{H i intensity mapping with MeerKAT: power spectrum detection in cross-correlation with WiggleZ galaxies}",
    journal = {\mnras},
    volume = {518},
    number = {4},
    pages = {6262-6272},
    year = {2022},
    month = {10},
    issn = {0035-8711},
    doi = {10.1093/mnras/stac3060},
    url = {https://doi.org/10.1093/mnras/stac3060},
    eprint = {https://academic.oup.com/mnras/article-pdf/518/4/6262/48302259/stac3060.pdf},
}

@article{Cunnington2025,
    author = {MeerKLASS Collaboration  and Bernal, José L and Bull, Philip and Camera, Stefano and Carucci, Isabella P and Chen, Zhaoting and Cunnington, Steven and Engelbrecht, Brandon N and Fonseca, José and Grainge, Keith and Irfan, Melis O and Li, Yichao and Mazumder, Aishrila and Paul, Sourabh and Pourtsidou, Alkistis and Santos, Mario G and Spinelli, Marta and Wang, Jingying and Witzemann, Amadeus and Wolz, Laura},
    title = {MeerKLASS L-band deep-field intensity maps: entering the H i dominated regime},
    journal = {Monthly Notices of the Royal Astronomical Society},
    volume = {537},
    number = {4},
    pages = {3632-3661},
    year = {2025},
    month = {02},
    issn = {0035-8711},
    doi = {10.1093/mnras/staf195},
    url = {https://doi.org/10.1093/mnras/staf195},
    eprint = {https://academic.oup.com/mnras/article-pdf/537/4/3632/61743817/staf195.pdf},
}

@misc{Cunnington2025b,
      title={Revealing cosmological fluctuations in 21cm intensity maps with MeerKLASS: from maps to power spectra}, 
      author={Steven Cunnington and Matilde Barberi-Squarotti and José Luis Bernal and Stefano Camera and Isabella P. Carucci and Zhaoting Chen and José Fonseca and Mario Santos and Marta Spinelli and Jingying Wang and Laura Wolz},
      year={2025},
      eprint={2510.27549},
      archivePrefix={arXiv},
      primaryClass={astro-ph.CO},
      url={https://arxiv.org/abs/2510.27549}, 
}

@incollection{Cunnington01.2026.SKA, author = {Steven Cunnington\&Wang and author2 and author3 and author4 and author5},title = {},year = {2026},publisher = {},note = {arXiv search: Report number AASKAII/Cunnington01},booktitle = {Advancing Astrophysics with the SKA -- II (AASKAII)}}

@article{Duchesne2023,
   title={The Rapid ASKAP Continuum Survey IV: continuum imaging at 1367.5 MHz and the first data release of RACS-mid},
   volume={40},
   ISSN={1448-6083},
   url={http://dx.doi.org/10.1017/pasa.2023.31},
   DOI={10.1017/pasa.2023.31},
   journal={Publications of the Astronomical Society of Australia},
   publisher={Cambridge University Press (CUP)},
   author={Duchesne, S. W. and Thomson, A. J. M. and Pritchard, J. and Lenc, E. and Moss, V. A. and McConnell, D. and Wieringa, M. H. and Whiting, M. T. and Wang, Z. and Wang, Y. and Rose, K. and Raja, W. and Murphy, Tara and Leung, J. K. and Huynh, M. T. and Hotan, A. W. and Hodgson, T. and Heald, G. H.},
   year={2023} }

@article{duchesne2025,
  title={The Rapid ASKAP Continuum Survey (RACS) VI: The RACS-high 1655.5 MHz images and catalogue},
  author={Duchesne, SW and Ross, K and Thomson, AJM and Lenc, E and Murphy, Tara and Galvin, TJ and Hotan, AW and Moss, VA and Whiting, Matthew T},
  journal={arXiv preprint arXiv:2501.04978},
  year={2025}}

@article{Engelbrecht2024,
    author = {Engelbrecht, Brandon N and Santos, Mario G and Fonseca, José and Li, Yichao and Wang, Jingying and Irfan, Melis O and Harper, Stuart E and Grainge, Keith and Bull, Philip and Carucci, Isabella P and Cunnington, Steven and Pourtsidou, Alkistis and Spinelli, Marta and Wolz, Laura},
    title = {Radio frequency interference from radio navigation satellite systems: simulations and comparison to MeerKAT single-dish data},
    journal = {Monthly Notices of the Royal Astronomical Society},
    volume = {536},
    number = {1},
    pages = {1035-1055},
    year = {2024},
    month = {11},
    issn = {0035-8711},
    doi = {10.1093/mnras/stae2649},
    url = {https://doi.org/10.1093/mnras/stae2649},
    eprint = {https://academic.oup.com/mnras/article-pdf/536/1/1035/61049770/stae2649.pdf},
}

@INPROCEEDINGS{Gupta2016,
       author = {{Gupta}, N. and {Srianand}, R. and {Baan}, W. and {Baker}, A.~J. and {Beswick}, R.~J. and {Bhatnagar}, S. and {Bhattacharya}, D. and {Bosma}, A. and {Carilli}, C. and {Cluver}, M. and {Combes}, F. and {Cress}, C. and {Dutta}, R. and {Fynbo}, J. and {Heald}, G. and {Hilton}, M. and {Hussain}, T. and {Jarvis}, M. and {Jozsa}, G. and {Kamphuis}, P. and {Kembhavi}, A. and {Kerp}, J. and {Kloeckner}, H.~R. and {Krogager}, J. and {Kulkarni}, V.~P. and {Ledoux}, C. and {Mahabal}, A. and {Mauch}, T. and {Moodley}, K. and {Momjian}, E. and {Morganti}, R. and {Noterdaeme}, P. and {Oosterloo}, T. and {Petitjean}, P. and {Schroeder}, A. and {Serra}, P. and {Sievers}, J. and {Spekkens}, K. and {Vaisanen}, P. and {van der Hulst}, T. and {Vivek}, M. and {Wang}, J. and {Wong}, O.~I. and {Zungu}, A.~R.},
        title = "{The MeerKAT Absorption Line Survey (MALS)}",
    booktitle = {MeerKAT Science: On the Pathway to the SKA},
         year = 2016,
        month = jan,
          eid = {14},
        pages = {14},
          doi = {10.22323/1.277.0014},
archivePrefix = {arXiv},
       eprint = {1708.07371},
 primaryClass = {astro-ph.GA},
       adsurl = {https://ui.adsabs.harvard.edu/abs/2016mks..confE..14G}
}

@article{Hale2021,
   title={The Rapid ASKAP Continuum Survey Paper II: First Stokes I Source Catalogue Data Release},
   volume={38},
   ISSN={1448-6083},
   url={http://dx.doi.org/10.1017/pasa.2021.47},
   DOI={10.1017/pasa.2021.47},
   journal={Publications of the Astronomical Society of Australia},
   publisher={Cambridge University Press (CUP)},
   author={Hale, Catherine L. and McConnell, D. and Thomson, A. J. M. and Lenc, E. and Heald, G. H. and Hotan, A. W. and Leung, J. K. and Moss, V. A. and Murphy, T. and Pritchard, J. and Sadler, E. M. and Stewart, A. J. and Whiting, M. T.},
   year={2021} }

@article{Hale2023,
    author = {Hale, C L and Schwarz, D J and Best, P N and Nakoneczny, S J and Alonso, D and Bacon, D and Böhme, L and Bhardwaj, N and Bilicki, M and Camera, S and Heneka, C S and Pashapour-Ahmadabadi, M and Tiwari, P and Zheng, J and Duncan, K J and Jarvis, M J and Kondapally, R and Magliocchetti, M and Rottgering, H J A and Shimwell, T W},
    title = {Cosmology from LOFAR Two-metre Sky Survey Data Release 2: angular clustering of radio sources},
    journal = {Monthly Notices of the Royal Astronomical Society},
    volume = {527},
    number = {3},
    pages = {6540-6568},
    year = {2023},
    month = {10},
    issn = {0035-8711},
    doi = {10.1093/mnras/stad3088},
    url = {https://doi.org/10.1093/mnras/stad3088},
    eprint = {https://academic.oup.com/mnras/article-pdf/527/3/6540/54082865/stad3088.pdf},
}

@incollection{Hale01.2026.SKA, author = {Catherine Hale and Fatemeh Tabatabaei},title = {},year = {2026},publisher = {},note = {arXiv search: Report number AASKAII/Hale01},booktitle = {Advancing Astrophysics with the SKA -- II (AASKAII)}}

@article{Harper2018,
    author = {Harper, Stuart E and Dickinson, Clive},
    title = {Potential impact of global navigation satellite services on total power H i intensity mapping surveys},
    journal = {Monthly Notices of the Royal Astronomical Society},
    volume = {479},
    number = {2},
    pages = {2024-2036},
    year = {2018},
    month = {06},
    issn = {0035-8711},
    doi = {10.1093/mnras/sty1495},
    url = {https://doi.org/10.1093/mnras/sty1495},
    eprint = {https://academic.oup.com/mnras/article-pdf/479/2/2024/25142146/sty1495.pdf},
}

@ARTICLE{Haslam1982,
   author = {{Haslam}, C.~G.~T. and {Salter}, C.~J. and {Stoffel}, H. and 
	{Wilson}, W.~E.},
    title = "{A 408 MHz all-sky continuum survey. II - The atlas of contour maps}",
  journal = {\aaps},
     year = 1982,
    month = jan,
   volume = 47,
    pages = {1},
   adsurl = {http://adsabs.harvard.edu/abs/1982A%26AS...47....1H}
}

@article{Hopkins_2025, 
title={The Evolutionary Map of the Universe: A new radio atlas for the southern hemisphere sky}, 
volume={42}, 
DOI={10.1017/pasa.2025.10042}, 
journal={Publications of the Astronomical Society of Australia}, 
author={Hopkins, Andrew and Kapinska, Anna and Marvil, Joshua and Vernstrom, Tessa and Collier, Jordan and Norris, Ray and Gordon, Yjan and Duchesne, Stefan and Rudnick, Lawrence and Gupta, Nikhel and et al.}, 
year={2025}, 
pages={e071}
}

@INPROCEEDINGS{Hugo2022,
       author = {{Hugo}, Benjamin V. and {Perkins}, S. and {Merry}, B. and {Mauch}, T. and {Smirnov}, O.~M.},
        title = "{Tricolour: An Optimized SumThreshold Flagger for MeerKAT}",
    booktitle = {Astronomical Data Analysis Software and Systems XXX},
         year = 2022,
       editor = {{Ruiz}, Jose Enrique and {Pierfedereci}, Francesco and {Teuben}, Peter},
       series = {Astronomical Society of the Pacific Conference Series},
       volume = {532},
        month = jul,
        pages = {541},
          doi = {10.48550/arXiv.2206.09179},
archivePrefix = {arXiv},
       eprint = {2206.09179},
 primaryClass = {astro-ph.IM},
       adsurl = {https://ui.adsabs.harvard.edu/abs/2022ASPC..532..541H}
}

@article{Hurley-Walker2017,
    author = {Hurley-Walker, N and Callingham, J R and Hancock, P J and Franzen, T M O and Hindson, L and Kapińska, A D and Morgan, J and Offringa, A R and Wayth, R B and Wu, C and Zheng, Q and Murphy, T and Bell, M E and Dwarakanath, K S and For, B and Gaensler, B M and Johnston-Hollitt, M and Lenc, E and Procopio, P and Staveley-Smith, L and Ekers, R and Bowman, J D and Briggs, F and Cappallo, R J and Deshpande, A A and Greenhill, L and Hazelton, B J and Kaplan, D L and Lonsdale, C J and McWhirter, S R and Mitchell, D A and Morales, M F and Morgan, E and Oberoi, D and Ord, S M and Prabu, T and Shankar, N Udaya and Srivani, K S and Subrahmanyan, R and Tingay, S J and Webster, R L and Williams, A and Williams, C L},
    title = "{GaLactic and Extragalactic All-sky Murchison Widefield Array (GLEAM) survey – I. A low-frequency extragalactic catalogue}",
    journal = {Monthly Notices of the Royal Astronomical Society},
    volume = {464},
    number = {1},
    pages = {1146-1167},
    year = {2016},
    month = {09},
    issn = {0035-8711},
    doi = {10.1093/mnras/stw2337},
    url = {https://doi.org/10.1093/mnras/stw2337},
    eprint = {https://academic.oup.com/mnras/article-pdf/464/1/1146/47736764/mnras\_464\_1\_1146.pdf},
}

@article{Intema2016,
	author = {Intema, H. T. and Jagannathan, P. and Mooley, K. P. and Frail, D. A.},
	title = {The GMRT 150 MHz all-sky radio survey - First alternative data release TGSS ADR1},
	DOI= "10.1051/0004-6361/201628536",
	url= "https://doi.org/10.1051/0004-6361/201628536",
	journal = {A\&A},
	year = 2017,
	volume = 598,
	pages = "A78",
}

@INPROCEEDINGS{Jonas2016,
       author = {{Jonas}, J. and {MeerKAT Team}},
        title = "{The MeerKAT Radio Telescope}",
    booktitle = {MeerKAT Science: On the Pathway to the SKA},
         year = 2016,
        month = jan,
          eid = {1},
        pages = {1},
          doi = {10.22323/1.277.0001},
       adsurl = {https://ui.adsabs.harvard.edu/abs/2016mks..confE...1J}
}

@ARTICLE{KiDS_DR5,
       author = {{Wright}, Angus H. and {Kuijken}, Konrad and {Hildebrandt}, Hendrik and {Radovich}, Mario and {Bilicki}, Maciej and {Dvornik}, Andrej and {Getman}, Fedor and {Heymans}, Catherine and {Hoekstra}, Henk and {Li}, Shun-Sheng and {Miller}, Lance and {Napolitano}, Nicola R. and {Xia}, Qianli and {Asgari}, Marika and {Brescia}, Massimo and {Buddelmeijer}, Hugo and {Burger}, Pierre and {Castignani}, Gianluca and {Cavuoti}, Stefano and {de Jong}, Jelte and {Edge}, Alastair and {Giblin}, Benjamin and {Giocoli}, Carlo and {Harnois-D{\'e}raps}, Joachim and {Jalan}, Priyanka and {Joachimi}, Benjamin and {John William}, Anjitha and {Joudaki}, Shahab and {Kannawadi}, Arun and {Kaur}, Gursharanjit and {La Barbera}, Francesco and {Linke}, Laila and {Mahony}, Constance and {Maturi}, Matteo and {Moscardini}, Lauro and {Nakoneczny}, Szymon J. and {Paolillo}, Maurizio and {Porth}, Lucas and {Puddu}, Emanuella and {Reischke}, Robert and {Schneider}, Peter and {Sereno}, Mauro and {Shan}, HuanYuan and {Sif{\'o}n}, Crist{\'o}bal and {St{\"o}lzner}, Benjamin and {Tr{\"o}ster}, Tilman and {Valentijn}, Edwin and {van den Busch}, Jan Luca and {Verdoes Kleijn}, Gijs and {Wittje}, Anna and {Yan}, Ziang and {Yao}, Ji and {Yoon}, Mijin and {Zhang}, Yun-Hao},
      journal = {AAP},
         year = 2024,
        month = jun,
       volume = {686},
          eid = {A170},
        pages = {A170},
          doi = {10.1051/0004-6361/202346730},
archivePrefix = {arXiv},
       eprint = {2503.19439},
 primaryClass = {astro-ph.GA},
       adsurl = {https://ui.adsabs.harvard.edu/abs/2024A&A...686A.170W}
}

@article{Lacy2020,
doi = {10.1088/1538-3873/ab63eb},
url = {https://dx.doi.org/10.1088/1538-3873/ab63eb},
year = {2020},
month = {jan},
publisher = {The Astronomical Society of the Pacific},
volume = {132},
number = {1009},
pages = {035001},
author = {Lacy, M. and Baum, S. A. and Chandler, C. J. and Chatterjee, S. and Clarke, T. E. and Deustua, S. and English, J. and Farnes, J. and Gaensler, B. M. and Gugliucci, N. and Hallinan, G. and Kent, B. R. and Kimball, A. and Law, C. J. and Lazio, T. J. W. and Marvil, J. and Mao, S. A. and Medlin, D. and Mooley, K. and Murphy, E. J. and Myers, S. and Osten, R. and Richards, G. T. and Rosolowsky, E. and Rudnick, L. and Schinzel, F. and Sivakoff, G. R. and Sjouwerman, L. O. and Taylor, R. and White, R. L. and Wrobel, J. and Andernach, H. and Beasley, A. J. and Berger, E. and Bhatnager, S. and Birkinshaw, M. and Bower, G. C. and Brandt, W. N. and Brown, S. and Burke-Spolaor, S. and Butler, B. J. and Comerford, J. and Demorest, P. B. and Fu, H. and Giacintucci, S. and Golap, K. and Güth, T. and Hales, C. A. and Hiriart, R. and Hodge, J. and Horesh, A. and Ivezić, Ž. and Jarvis, M. J. and Kamble, A. and Kassim, N. and Liu, X. and Loinard, L. and Lyons, D. K. and Masters, J. and Mezcua, M. and Moellenbrock, G. A. and Mroczkowski, T. and Nyland, K. and O’Dea, C. P. and O’Sullivan, S. P. and Peters, W. M. and Radford, K. and Rao, U. and Robnett, J. and Salcido, J. and Shen, Y. and Sobotka, A. and Witz, S. and Vaccari, M. and Weeren, R. J. van and Vargas, A. and Williams, P. K. G. and Yoon, I.},
title = {The Karl G. Jansky Very Large Array Sky Survey (VLASS). Science Case and Survey Design},
journal = {Publications of the Astronomical Society of the Pacific}
}

@article{LOLSS_2023,
	author = {{de Gasperin, F.} and {Edler, H. W.} and {Williams, W. L.} and {Callingham, J. R.} and {Asabere, B.} and {Brüggen, M.} and {Brunetti, G.} and {Dijkema, T. J.} and {Hardcastle, M. J.} and {Iacobelli, M.} and {Offringa, A.} and {Norden, M. J.} and {Röttgering, H. J. A.} and {Shimwell, T.} and {van Weeren, R. J.} and {Tasse, C.} and {Bomans, D. J.} and {Bonafede, A.} and {Botteon, A.} and {Cassano, R.} and {Chyży, K. T.} and {Cuciti, V.} and {Emig, K. L.} and {Kadler, M.} and {Miley, G.} and {Mingo, B.} and {Oei, M. S. S. L.} and {Prandoni, I.} and {Schwarz, D. J.} and {Zarka, P.}},
	title = {The LOFAR LBA Sky Survey - II. First data release★},
	DOI= "10.1051/0004-6361/202245389",
	url= "https://doi.org/10.1051/0004-6361/202245389",
	journal = {A\&A},
	year = 2023,
	volume = 673,
	pages = "A165",
}

@ARTICLE{Mangum2007,
       author = {{Mangum}, J.~G. and {Emerson}, D.~T. and {Greisen}, E.~W.},
        title = "{The On The Fly imaging technique}",
      journal = {\aap},
         year = 2007,
        month = nov,
       volume = {474},
       number = {2},
        pages = {679-687},
          doi = {10.1051/0004-6361:20077811},
archivePrefix = {arXiv},
       eprint = {0709.0553},
 primaryClass = {astro-ph},
       adsurl = {https://ui.adsabs.harvard.edu/abs/2007A&A...474..679M}
}

@article{Mauch2003,
    author = {Mauch, T. and Murphy, T. and Buttery, H. J. and Curran, J. and Hunstead, R. W. and Piestrzynski, B. and Robertson, J. G. and Sadler, E. M.},
    title = {SUMSS: a wide-field radio imaging survey of the southern sky – II. The source catalogue},
    journal = {Monthly Notices of the Royal Astronomical Society},
    volume = {342},
    number = {4},
    pages = {1117-1130},
    year = {2003},
    month = {07},
    issn = {0035-8711},
    doi = {10.1046/j.1365-8711.2003.06605.x},
    url = {https://doi.org/10.1046/j.1365-8711.2003.06605.x},
    eprint = {https://academic.oup.com/mnras/article-pdf/342/4/1117/2819646/342-4-1117.pdf},
}

@article{McConnell2020,
   title={The Rapid ASKAP Continuum Survey I: Design and first results},
   volume={37},
   ISSN={1448-6083},
   url={http://dx.doi.org/10.1017/pasa.2020.41},
   DOI={10.1017/pasa.2020.41},
   journal={Publications of the Astronomical Society of Australia},
   publisher={Cambridge University Press (CUP)},
   author={McConnell, D. and Hale, C. L. and Lenc, E. and Banfield, J. K. and Heald, George and Hotan, A. W. and Leung, James K. and Moss, Vanessa A. and Murphy, Tara and O’Brien, Andrew and Pritchard, Joshua and Raja, Wasim and Sadler, Elaine M. and Stewart, Adam and Thomson, Alec J. M. and Whiting, M. and Allison, James R. and Amy, S. W. and Anderson, C. and Ball, Lewis and Bannister, Keith W. and Bell, Martin and Bock, Douglas C.-J. and Bolton, Russ and Bunton, J. D. and Chippendale, A. P. and Collier, J. D. and Cooray, F. R. and Cornwell, T. J. and Diamond, P. J. and Edwards, P. G. and Gupta, N. and Hayman, Douglas B. and Heywood, Ian and Jackson, C. A. and Koribalski, Bärbel S. and Lee-Waddell, Karen and McClure-Griffiths, N. M. and Ng, Alan and Norris, Ray P. and Phillips, Chris and Reynolds, John E. and Roxby, Daniel N. and Schinckel, Antony E. T. and Shields, Matt and Tremblay, Chenoa and Tzioumis, A. and Voronkov, M. A. and Westmeier, Tobias},
   year={2020} }

@article{Mooley2019,
doi = {10.3847/1538-4357/aaef7c},
url = {https://dx.doi.org/10.3847/1538-4357/aaef7c},
year = {2018},
month = {dec},
publisher = {The American Astronomical Society},
volume = {870},
number = {1},
pages = {25},
author = {Mooley, K. P. and Myers, S. T. and Frail, D. A. and Hallinan, G. and Butler, B. and Kimball, A. and Golap, K.},
title = {The Caltech-NRAO Stripe 82 Survey (CNSS). II. On-the-fly Mosaicking Methodology},
journal = {The Astrophysical Journal}
}

@article{Murphy2010,
    author = {Murphy, Tara and Sadler, Elaine M. and Ekers, Ronald D. and Massardi, Marcella and Hancock, Paul J. and Mahony, Elizabeth and Ricci, Roberto and Burke-Spolaor, Sarah and Calabretta, Mark and Chhetri, Rajan and De Zotti, Gianfranco and Edwards, Philip G. and Ekers, Jennifer A. and Jackson, Carole A. and Kesteven, Michael J. and Lindley, Emma and Newton-McGee, Katherine and Phillips, Chris and Roberts, Paul and Sault, Robert J. and Staveley-Smith, Lister and Subrahmanyan, Ravi and Walker, Mark A. and Wilson, Warwick E.},
    title = {The Australia Telescope 20 GHz Survey: the source catalogue},
    journal = {Monthly Notices of the Royal Astronomical Society},
    volume = {402},
    number = {4},
    pages = {2403-2423},
    year = {2010},
    month = {03},
    issn = {0035-8711},
    doi = {10.1111/j.1365-2966.2009.15961.x},
    url = {https://doi.org/10.1111/j.1365-2966.2009.15961.x},
    eprint = {https://academic.oup.com/mnras/article-pdf/402/4/2403/17323327/mnras0402-2403.pdf},
}

@article{Norris2011, 
title={EMU: Evolutionary Map of the Universe}, 
volume={28}, 
DOI={10.1071/AS11021}, 
number={3}, 
journal={Publications of the Astronomical Society of Australia}, 
author={Norris, Ray P. and Hopkins, A. M. and Afonso, J. and Brown, S. and Condon, J. J. and Dunne, L. and Feain, I. and Hollow, R. and Jarvis, M. and Johnston-Hollitt, M. and et al.}, 
year={2011}, 
pages={215–248}
}

@ARTICLE{Offringa2010,
       author = {{Offringa}, A.~R. and {de Bruyn}, A.~G. and {Biehl}, M. and {Zaroubi}, S. and {Bernardi}, G. and {Pandey}, V.~N.},
        title = "{Post-correlation radio frequency interference classification methods}",
      journal = {\mnras},
         year = 2010,
        month = jun,
       volume = {405},
       number = {1},
        pages = {155-167},
          doi = {10.1111/j.1365-2966.2010.16471.x},
archivePrefix = {arXiv},
       eprint = {1002.1957},
 primaryClass = {astro-ph.IM},
       adsurl = {https://ui.adsabs.harvard.edu/abs/2010MNRAS.405..155O}
}

@inproceedings{Santos2017,
      author         = "Santos, Mario G. and others",
      title          = "{MeerKLASS: MeerKAT Large Area Synoptic Survey}",
      booktitle      = "{Proceedings, MeerKAT Science: On the Pathway to the SKA
                        (MeerKAT2016): Stellenbosch, South Africa, May 25-27,
                        2016}",
      collaboration  = "MeerKLASS",
      year           = "2017",
      eprint         = "1709.06099",
      archivePrefix  = "arXiv",
      primaryClass   = "astro-ph.CO",
      SLACcitation   = "%%CITATION = ARXIV:1709.06099;%%"
}

@ARTICLE{Sawada2008,
       author = {{Sawada}, Tsuyoshi and {Ikeda}, Norio and {Sunada}, Kazuyoshi and {Kuno}, Nario and {Kamazaki}, Takeshi and {Morita}, Koh-Ichiro and {Kurono}, Yasutaka and {Koura}, Norikazu and {Abe}, Katsumi and {Kawase}, Sachiko and {Maekawa}, Jun and {Horigome}, Osamu and {Yanagisawa}, Kiyohiko},
        title = "{On-The-Fly Observing System of the Nobeyama 45-m and ASTE 10-m Telescopes}",
      journal = {\pasj},
         year = 2008,
        month = jun,
       volume = {60},
        pages = {445},
          doi = {10.1093/pasj/60.3.445},
archivePrefix = {arXiv},
       eprint = {0712.1283},
 primaryClass = {astro-ph},
       adsurl = {https://ui.adsabs.harvard.edu/abs/2008PASJ...60..445S}
}

@article{Shimwell2022,
	author = {{Shimwell, T. W.} and {Hardcastle, M. J.} and {Tasse, C.} and {Best, P. N.} and {R\"ottgering, H. J. A.} and {Williams, W. L.} and {Botteon, A.} and {Drabent, A.} and {Mechev, A.} and {Shulevski, A.} and {van Weeren, R. J.} and {Bester, L.} and {Br\"uggen, M.} and {Brunetti, G.} and {Callingham, J. R.} and {Chyzy, K. T.} and {Conway, J. E.} and {Dijkema, T. J.} and {Duncan, K.} and {de Gasperin, F.} and {Hale, C. L.} and {Haverkorn, M.} and {Hugo, B.} and {Jackson, N.} and {Mevius, M.} and {Miley, G. K.} and {Morabito, L. K.} and {Morganti, R.} and {Offringa, A.} and {Oonk, J. B. R.} and {Rafferty, D.} and {Sabater, J.} and {Smith, D. J. B.} and {Schwarz, D. J.} and {Smirnov, O.} and {O\'{}Sullivan, S. P.} and {Vedantham, H.} and {White, G. J.} and {Albert, J. G.} and {Alegre, L.} and {Asabere, B.} and {Bacon, D. J.} and {Bonafede, A.} and {Bonnassieux, E.} and {Brienza, M.} and {Bilicki, M.} and {Bonato, M.} and {Calistro Rivera, G.} and {Cassano, R.} and {Cochrane, R.} and {Croston, J. H.} and {Cuciti, V.} and {Dallacasa, D.} and {Danezi, A.} and {Dettmar, R. J.} and {Di Gennaro, G.} and {Edler, H. W.} and {En\ss{}lin, T. A.} and {Emig, K. L.} and {Franzen, T. M. O.} and {Garc\'{\i}a-Vergara, C.} and {Grange, Y. G.} and {G\"urkan, G.} and {Hajduk, M.} and {Heald, G.} and {Heesen, V.} and {Hoang, D. N.} and {Hoeft, M.} and {Horellou, C.} and {Iacobelli, M.} and {Jamrozy, M.} and {Jeli\'{}c, V.} and {Kondapally, R.} and {Kukreti, P.} and {Kunert-Bajraszewska, M.} and {Magliocchetti, M.} and {Mahatma, V.} and {Malek, K.} and {Mandal, S.} and {Massaro, F.} and {Meyer-Zhao, Z.} and {Mingo, B.} and {Mostert, R. I. J.} and {Nair, D. G.} and {Nakoneczny, S. J.} and {Nikiel-Wroczy\'{}nski, B.} and {Orr\'u, E.} and {Pajdosz-\'{}Smierciak, U.} and {Pasini, T.} and {Prandoni, I.} and {van Piggelen, H. E.} and {Rajpurohit, K.} and {Retana-Montenegro, E.} and {Riseley, C. J.} and {Rowlinson, A.} and {Saxena, A.} and {Schrijvers, C.} and {Sweijen, F.} and {Siewert, T. M.} and {Timmerman, R.} and {Vaccari, M.} and {Vink, J.} and {West, J. L.} and {Wolowska, A.} and {Zhang, X.} and {Zheng, J.}},
	title = {The LOFAR Two-metre Sky Survey - V. Second data release},
	DOI= "10.1051/0004-6361/202142484",
	url= "https://doi.org/10.1051/0004-6361/202142484",
	journal = {A\&A},
	year = 2022,
	volume = 659,
	pages = "A1",
}

@ARTICLE{Smirnov2011,
   author = {{Smirnov}, O.~M.},
    title = "{Revisiting the radio interferometer measurement equation. II. Calibration and direction-dependent effects}",
  journal = {\aap},
archivePrefix = "arXiv",
   eprint = {1101.1765},
 primaryClass = "astro-ph.IM",
     year = 2011,
    month = mar,
   volume = 527,
      eid = {A107},
    pages = {A107},
      doi = {10.1051/0004-6361/201116434},
   adsurl = {http://adsabs.harvard.edu/abs/2011A%26A...527A.107S}
}

@incollection{Spinelli02.2026.SKA, author = {Ian Harrison and author2 and author3 and author4 and author5},title = {},year = {2026},publisher = {},note = {arXiv search: Report number AASKAII/Spinelli02},booktitle = {Advancing Astrophysics with the SKA -- II (AASKAII)}}

@article{Tasse2018,
   title={Faceting for direction-dependent spectral deconvolution},
   volume={611},
   ISSN={1432-0746},
   url={http://dx.doi.org/10.1051/0004-6361/201731474},
   DOI={10.1051/0004-6361/201731474},
   journal={Astronomy \& Astrophysics},
   publisher={EDP Sciences},
   author={Tasse, C. and Hugo, B. and Mirmont, M. and Smirnov, O. and Atemkeng, M. and Bester, L. and Hardcastle, M. J. and Lakhoo, R. and Perkins, S. and Shimwell, T.},
   year={2018},
   month=mar, pages={A87} }

@article{Wang2021,
    author = {Wang, Jingying and Santos, Mario G and Bull, Philip and Grainge, Keith and Cunnington, Steven and Fonseca, José and Irfan, Melis O and Li, Yichao and Pourtsidou, Alkistis and Soares, Paula S and Spinelli, Marta and Bernardi, Gianni and Engelbrecht, Brandon},
    title = {H i intensity mapping with MeerKAT: calibration pipeline for multidish autocorrelation observations},
    journal = {Monthly Notices of the Royal Astronomical Society},
    volume = {505},
    number = {3},
    pages = {3698-3721},
    year = {2021},
    month = {05},
    issn = {0035-8711},
    doi = {10.1093/mnras/stab1365},
    url = {https://doi.org/10.1093/mnras/stab1365},
    eprint = {https://academic.oup.com/mnras/article-pdf/505/3/3698/38711683/stab1365.pdf},
}

@article{ Wagenveld2024,
	author = {{Wagenveld, J. D.} and {Klöckner, H.-R.} and {Gupta, N.} and {Sekhar, S.} and {Jagannathan, P.} and {Deka, P. P.} and {Jose, J.} and {Balashev, S. A.} and {Borgaonkar, D.} and {Chatterjee, A.} and {Combes, F.} and {Emig, K. L.} and {Gaunekar, A. N.} and {Hilton, M.} and {Józsa, G. I. G.} and {Klutse, D. Y.} and {Knowles, K.} and {Krogager, J.-K.} and {Momjian, E.} and {Muller, S.} and {Sikhosana, S. P.}},
	title = {The MeerKAT Absorption Line Survey Data Release 2: Wideband continuum catalogues and a measurement of the cosmic radio dipole★},
	DOI= "10.1051/0004-6361/202450291",
	url= "https://doi.org/10.1051/0004-6361/202450291",
	journal = {A\&A},
	year = 2024,
	volume = 690,
	pages = "A163",
}

@article{Wayth2015, 
title={GLEAM: The GaLactic and Extragalactic All-Sky MWA Survey}, 
volume={32}, 
DOI={10.1017/pasa.2015.26}, 
journal={Publications of the Astronomical Society of Australia}, 
author={Wayth, R. B. and Lenc, E. and Bell, M. E. and Callingham, J. R. and Dwarakanath, K. S. and Franzen, T. M. O. and For, B.-Q. and Gaensler, B. and Hancock, P. and Hindson, L. and et al.}, 
year={2015}, 
pages={e025}
}

\end{document}